\documentclass[pre,showpacs,twocolumn,superscriptaddress,floatfix,amsmath]{revtex4}
 \usepackage[latin1]{inputenc}
 \usepackage[english]{babel}
 \usepackage{graphicx}
 \usepackage{psfrag}
 \usepackage{amssymb}
 \usepackage{amsmath}
 \usepackage{amscd}
 \usepackage{eucal}
 \usepackage{color}
 \usepackage{bm}
 
 \newcommand{\ignore}[1]{\relax}

\newcommand{\de}{\delta}

\newcommand{\eps}{\epsilon}

\newcommand{\tEc}{\tau_{\rm E}^{\rm c}}
\newcommand{\tEo}{\tau_{\rm E}^{\rm o}}
\newcommand{\tEd}{\tau_{\rm E}^{\rm d}}
\newcommand{\tD}{\tau_{\rm D}}

 \newcommand{\e}{{\rm e}}
 \newcommand{\rmd}{{\rm d}}
 \newcommand{\rmi}{{\rm i}}
 
 \newcommand{\tr}{{\rm tr}}

\begin{document}

\title{Semiclassical transport in nearly symmetric quantum dots I: \\
symmetry-breaking in the dot}

\author{Robert S. Whitney}
\affiliation{Institut Laue-Langevin, 6 rue Jules Horowitz, B.P. 156,
         38042 Grenoble, France}

\author{Henning Schomerus}
\author{Marten Kopp }
\affiliation{Department of Physics, Lancaster University, Lancaster, LA1 4YB, United Kingdom}

\date{June 4, 2009}

\begin{abstract}
We apply the semiclassical theory of transport to quantum dots
with exact and approximate spatial symmetries; left-right mirror
symmetry, up-down mirror symmetry, inversion symmetry or four-fold
symmetry. In this work  --- the first of a pair of articles --- we
consider (a) perfectly symmetric dots and (b) nearly symmetric
dots in which the symmetry is broken by the dot's internal
dynamics. The second article addresses symmetry-breaking by
displacement of the leads. Using semiclassics, we identify the
origin of the symmetry-induced interference effects that
contribute to weak localization corrections and universal
conductance fluctuations.
For perfect spatial symmetry, we recover results previously
found using the random-matrix theory conjecture. We then go on to 
show how the
results are affected by asymmetries in the dot, magnetic fields
and decoherence. In particular, the symmetry-asymmetry crossover
is found to be described by a universal dependence on an asymmetry
parameter $\gamma_{\rm asym}$. However, the form of this parameter
is very different depending on how the dot is deformed away from
spatial symmetry. 
Symmetry-induced interference effects are completely destroyed
when the dot's boundary is globally deformed by less than an electron 
wavelength. In contrast, these effects are only reduced
by a finite amount when a part of the dot's boundary smaller than
a lead-width is deformed an arbitrarily large distance.
\end{abstract}

\pacs{05.45.Mt,74.40.+k,73.23.-b,03.65.Yz}

\maketitle


\section{Introduction}

In recent years there has been a great deal of progress in using
semiclassical methods to describe the quantum behaviour of systems
which are classically chaotic. Semiclassics was applied to
analyzing the transport properties of chaotic quantum dots in
Refs.~\cite{Bar91,Bar93}. Despite clear indications from the
quasiclassical method
\cite{Larkin-Khem,Cha86,Aleiner-Larkin,Agam}, it took many years
to understand how to go beyond the diagonal approximation
\cite{Berry,Sie01,Haake-rmt}, and thereby use semiclassics to
analyze mesoscopic effects such as weak localization
\cite{Richter-Sieber,
haake-conductance,Jac06,Rahav-Brouwer-backscatter}, universal
conductance fluctuations
\cite{Brouwer-Rahav-ucfs,Brouwer-Rahav-ucfs2} and shot-noise
\cite{Haake-fano,wj-fano,Haake-longpaper}. Before these works,
general statements on the quantum behaviour of such systems relied
on the Bohigas-Giannoni-Schmit conjecture
\cite{Bohigas-Giannoni-Schmit} that they are well-described by
random-matrix theory (RMT) \cite{Guhr,Beenakker-1997}. RMT is a
statistical theory based on the assumption that we know nothing
about the dynamics of the system (besides global symmetries such
as time-reversal and spin-rotation symmetry). However, often this
is not the case; perhaps the system has extra constants of motion
or the classical phase space of the system is mixed (so that some
parts are regular and others are chaotic). Such systems are hard
to treat using RMT methods. In those cases where progress with RMT
is possible, one adds more phenomenological parameters to the
theory, thereby reducing its predictive power. One of the great
hopes for semiclassics is that it will provide {\it microscopic}
theories for such situations, in which all parameters can be
derived from properties of the confining potential (and any
inter-particle interactions that are present).

A simpler, but related, problem is that of chaotic systems with
additional spatial symmetries (reflection or inversion
symmetries). One can ask two questions. Do spatial symmetries lead
to interference effects similar to those induced by time-reversal
symmetry (coherent backscattering and weak localization effects)?
If so, how are these interference effects changed by the presence
of spatial asymmetries or magnetic fields?

These discrete spatial symmetries are simple enough to be amenable
to a phenomenological analysis with RMT. Such an analysis was
carried out in Refs.~\cite{Baranger-Mello,Gopar96} where it was
shown that spatial symmetries do indeed lead to extra interference
effects analogous to weak localization. Since that work, there
have been a number of other studies of chaotic systems with
symmetries. Refs.\
\cite{Martinez-Mello,Kopp-Schomerus-Rotter,Gopar-Rotter-Schomerus}
use RMT methods to study the distribution of transmission
eigenvalues, while Ref.~\cite{WMM} identifies a huge
interference effect in double-dots separated by a tunnel barrier.
However, none of these works have addressed how the
weak localization correction and conductance fluctuations behave
for dots that are only approximately symmetric.

This is the first of a pair of articles in which we perform a
semiclassical analysis of symmetry-induced interference effects on
quantum transport, and in particular their contributions to
weak localization corrections and conductance fluctuations. 
We consider a quantum dot (billiard) whose shape has spatial symmetries (mirror or inversion symmetries) but is otherwise shaped such that the classical dynamics within it would exhibit hyperbolic chaos (exponential divergence of initially close trajectories).
We develop a semiclassical theory of transport between two ideal leads coupled to the dot,
building on the theory for similar chaotic systems without spatial symmetries. 
First, we give a non-phenomenological
explanation of the origin of the symmetry-induced interference
effects found in RMT \cite{Baranger-Mello,Gopar96}. We then
investigate how transport is affected by asymmetries in the dot,
magnetic fields, dephasing, and the Ehrenfest time (which
characterizes the emergence of fully developed wave chaos
\cite{Aleiner-Larkin,Schomerus-Jacquod}). For this we make use of
analogies with the analysis of dephasing in asymmetric dots
\cite{Bro06-quasi-ucf,pjw-decoh,wjp-decoh}.

We find that different types of
deformations away from symmetry destroy the symmetry-induced
interference effects in remarkably different ways. 
The deformations we consider have the same
universal dependence on an asymmetry parameter $\gamma_{\rm
asym}$; however, the microscopic form of this parameter varies
depending on the type of deformation. For a global deformation of
the dot (say half the dot is displaced outward), we find that a
displacement $\de L$ of less than a Fermi wavelength is sufficient
to completely destroy the symmetry-induced interference effects.
In contrast, for the deformation of a small part of the boundary
(say a portion less than but of order a lead-width is displaced
outwards) we find that an arbitrary large displacement $\de L$
does not destroy the symmetry-induced interference effects
completely, instead it simply suppresses them by a finite amount.

Finally, we address how asymmetries and magnetic fields can be incorporated
into a simple phenomenological RMT model, and compare results of numerical
computations in this model with the semiclassical predictions.

In the second of this pair of articles \cite{whitney-schomerus-kopp-2nd}, we
use the same methods to find these transport properties for a spatially
symmetric system coupled to leads which do not respect the symmetry.

\section{Scope and overview of this work}

 \begin{table}
 \hfill
 \begin{tabular}{|l|c c|c c|}
 \hline
 & \multicolumn{2}{c|}{\ Average conductance\ } &
\multicolumn{2}{c|}{\ UCFs\ }
  \\
&  $B\ll B_{\rm c}$ & $B\gg B_{\rm c}$ &  $B\ll B_{\rm c}$ \ & $B\gg B_{\rm c}$ \\
\hline
 \ no spatial sym.$\phantom{ {\displaystyle{1\atop 1}}}$&
${N\over 2}-{1 \over 4} $ &
${N \over 2}$ &  ${1 \over 8}$ & ${1 \over 16}$
\\ [1ex]
 \ left-right sym.\  &
${N\over 2}$ &
${N\over 2}+{1 \over 4}$
&  ${1 \over 4}$ & ${1 \over 8}$
\\ [1ex]
 \ inversion sym.\ &
${N\over 2}$ &  ${N\over 2}$
&  ${1 \over 4}$ & ${1 \over 8}$
\\[1ex]
 \ up-down sym.\ &
${N\over 2}-{1 \over 2}$ &  ${N\over 2}-{1 \over 4}$
&  ${1 \over 4}$ & ${1 \over 8}$
\\[1ex]
 \ four-fold sym.\ &
${N\over 2}$ &
${N\over 2}$
&  ${1 \over 2}$ & ${1 \over 4}$
\\[1ex]
 \hline
 \end{tabular}
%
%
 \caption{\label{table1}
Table of the values of average conductance and conductance
fluctuations for quantum dots with different spatial symmetries,
attached to two leads both carrying $N\gg 1$ modes. $B_{\rm c}$ is a
magnetic field strength at which time-reversal symmetry is broken
in the asymmetric situation. These results are taken from
Refs.~\cite{Baranger-Mello,Gopar96}, where they were derived in
RMT.  Here we rederive these results semiclassically, and show how
they depend on asymmetries in the dot, dephasing, magnetic fields
and Ehrenfest times. }
 \end{table}

In this article we consider two-dimensional chaotic quantum dots of
characteristic size $L$, perfectly coupled to two leads, labeled the left (L)
and right (R) lead. The leads have widths $W_{\rm L}$ and $W_{\rm R}$ much
greater than the Fermi wavelength $\lambda_{\rm F} = h/p_{\rm F}$. Thus  the
leads carry $N_{\rm L}$ and $N_{\rm R}$ modes, where $N_\kappa = p_{\rm
F}W_\kappa/(\pi\hbar) \gg 1$ for $\kappa \in {\rm L,R}$.

Assuming that ergodicity of the classical dynamics is established
quickly, there are only a few time scales which fully characterize
the quantum transport through the dot. The typical time that a
particle spends in the dot is given by the dwell time $\tD =
\tau_0 \times C/(W_{\rm L}+W_{\rm R})$, where $C$ is the dot's
circumference and $\tau_0=\pi A/C v_F$ is the scattering time off
the boundary (here $A$ is the area of the dot and $v_F$ is the
Fermi velocity).

It is now clear that the development of quantum interference
effects in such systems is governed by Ehrenfest times $\tau_{\rm
E}$ \cite{Aleiner-Larkin}. Semiclassics enables us to treat finite
Ehrenfest times, thereby exploring the crossover between RMT
behaviour ($\tau_{\rm E} \ll \tD$) and classical behaviour
($\tau_{\rm E} \gg \tD$). There is in fact a set of Ehrenfest
times, which all take the form \cite{Schomerus-Jacquod}
\begin{eqnarray}
\tau_{\rm E} = \Lambda^{-1} \ln [ (L/\lambda_{\rm F}) {\cal L}]
\label{Eq:tau_E}
\end{eqnarray}
where $\Lambda$ is the Lyapunov exponent of the classical dynamics
in the dot, and ${\cal L}$ is the ratio of characteristic
classical length scales. Of relevance for this paper are the
open-system Ehrenfest time $\tEo$, given by Eq.~(\ref{Eq:tau_E})
with ${\cal L}=(W/L)^2$ \cite{Vavilov-Larkin}, and the
closed-system Ehrenfest time $\tEc$  with ${\cal L}=1$ (a third
Ehrenfest time $\tEd$ with ${\cal L}=W/L$ is  relevant for decay
problems \cite{Schomerus-Tworzydlo}).

Two other relevant time scales are the
dephasing time $\tau_\phi=1/\gamma_\phi$ (where $\gamma_\phi$ is
the dephasing rate), and the time scale $\tau_B\sim(B_{\rm
0}/B)^2\tau_0$ on which a magnetic field destroys time-reversal
symmetry in the internal dynamics. Here, $B_{\rm 0} \sim h/(eA)$
is a characteristic field strength at which about one flux quantum
penetrates the quantum dot.  For transport, the effect of a
magnetic field is significant when it is of order
\begin{equation}
B_{\rm c}= a B_0\sqrt{\tau_0/2\tD}\label{eq:bc},
\end{equation}
where $a$ is a system-specific parameter  of order one
\cite{Beenakker-1997}.

There are three spatial symmetries of particular interest for
transport through a chaotic quantum dot coupled to two leads.  The
first two  map the L lead onto the R lead and vice versa, they are
a left-right mirror symmetry (Fig.~\ref{Fig:leftright}) and an
inversion symmetry (Fig.~\ref{Fig:inversion}). The third type of
symmetry maps each lead onto itself, such as an up-down mirror
symmetry (Fig.~\ref{Fig:updown}). A dot with any two of the three
symmetries listed above is four-fold symmetric, and automatically
satisfies the third symmetry. Four-fold symmetry is most easily
visualized by considering a dot which has both left-right and
up-down mirror symmetry. For dots with two leads there are no
additional cases of multiple symmetry.

Readers already familiar with the semiclassical theory of weak
localization \cite{Richter-Sieber} and coherent backscattering
 \cite{Jac06,Rahav-Brouwer-backscatter} may only need to
look at Figs.~\ref{Fig:leftright}-\ref{Fig:updown} to understand
the extra symmetry-induced interference effects. Once one folds
the paths shown under the relevant symmetry (the right-hand
sketches in each figure), we obtain contributions which are
similar to the  weak localization and coherent backscattering
contributions in a system with no spatial symmetries. Indeed,
given these figures, many such readers will probably be able to
correctly guess the form of the results for perfect symmetry,
which agree with the results of RMT tabulated in Table
\ref{table1}. Treating asymmetries, however, is more involved.

The rest of this article is laid out as follows. In
Section~\ref{sect:semiclass} we review the semiclassical
description of quantum transport, and discuss how it should be
modified to include spatial symmetries. In Section~\ref{sect:LR}
we treat weak localization in dots with left-right symmetry and
discuss the role of perpendicular magnetic fields, dephasing, and asymmetry in
the dot. In Sections \ref{sect:inversion}, \ref{sect:UD} and
\ref{sect:4fold} we analyze weak localization in dots with
inversion symmetry, up-down symmetry and four-fold symmetry,
respectively. 

In Section~\ref{sect:UCFs} we provide a semiclassical description
of conductance fluctuations in symmetric systems and discuss the
effects of asymmetries and perpendicular magnetic fields. To keep
the discussion simple we do not address dephasing here (unlike for
weak localization).

In Section~\ref{sect:RMT} we propose an efficient
phenomenological model to study symmetry breaking in RMT and
compare the results of numerical computations to the semiclassical
predictions.

Finally in Section~\ref{sect:expt}, we consider the application of our results to 
experimental systems. We estimate the necessary conditions for observing 
these symmetry-induced effects in the transport of electrons through quantum dots,
and of microwaves through chaotic cavities.

\begin{figure*}
\centerline{\hbox{\includegraphics[width=0.7\textwidth]{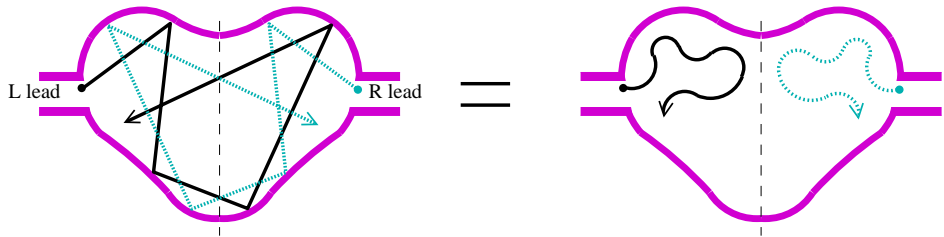}}}
\caption[]{\label{Fig:example} 
(colour online). Sketch of a planar quantum dot with
a left-right mirror symmetry. On the left we show a classical path
(solid line) and its mirror image (dashed line). Each path is
piecewise straight and undergoes specular reflections at the
boundary of the dot. The path is chaotic (we assume hyperbolic
chaos), so its position and momentum after time $t$ are
exponentially sensitive to the initial position and momentum. We
assume the system is such that a typical path crosses the symmetry
axis a large number of times before escaping into a lead (these
crossings occur on a timescale of order of the time-of-flight
across the system). To simplify the sketches elsewhere in this
article we draw paths as abstract curves (shown on the right)
which only show those crossings of the symmetry axis which are
crucial to the nature of the path.}
\end{figure*}

\section{The theory of semiclassical transport}
\label{sect:semiclass} The semiclassical picture of weak
localization in disordered systems has been known for many years
for both s-wave disorder \cite{Larkin-Khem,Cha86} and smooth
disorder \cite{Aleiner-Larkin}. However, only recently has it been
possible to extend this picture to disorder-free but chaotic
systems, such as ballistic quantum dots \cite{Richter-Sieber}. We
start with the semiclassical theory of transport
\cite{Bar91,Bar93} and sum over lead modes as in Ref.~\cite{Jac06}
(see also Appendix B in Ref.~\cite{Whi07}). Then the dimensionless
conductance (conductance in units of $2e^2/h$) is given by
\begin{eqnarray}\label{Eq:g-doublesum}
 g &=& {1 \over 2 \pi \hbar}
  \int_{\rm L}\!\!{\rm d} y_{0}   \int_{\rm R} \!{\rm d} y
    \sum_{ \gamma, \gamma'}
     A_{\gamma} A_{\gamma'} \;\;
    e^{ i (S_\gamma- S_{\gamma'})/\hbar},
 \end{eqnarray}
where the double sum is over all classical paths $\gamma,\gamma'$, from point
$y_0$ on lead L to point $y$ on lead R. Path $\gamma$ has a classical action
$S_\gamma= \int_\gamma {\bf p} \rmd {\bf r}$, and an amplitude $A_\gamma$.
Refs.~\cite{Bar91,Bar93} have shown that the squared-amplitude $A_\gamma^2
= (\rmd p_{y0}/\rmd y)$,
so it measures the path's stability 
(the change in the path's initial momentum $p_{y0}$
associated with a small displacement of its final position $y$).

For most pairs of $\gamma$ and $\gamma'$ the exponential in
Eq.~(\ref{Eq:g-doublesum}) oscillates wildly as one changes the
energy or the dot-shape. Thus they make no contribution to the
average conductance (where one averages over energy, dot-shape, or
both). The only contributions that survive averaging are those
where the pairs of paths have highly correlated actions; typically
one has $S_\gamma \simeq S_{\gamma'}$ for a broad range of
energies and dot-shapes. The simplest example of this are all
contributions to the double-sum with $\gamma'=\gamma$. These
``diagonal contributions'' to the above double sum can be
evaluated with the help of the sum rule (in the spirit of Eq.~(B6) of  Ref.~\cite{Bar91})
  \begin{eqnarray}
 \sum_{\gamma} A_{\gamma}^2 \left[ \cdots \right]_{\gamma}  \! &=& \! \!
 \int_{-\frac{\pi}{2}}^{\frac{\pi}{2}} {\rm d} \theta_{0}
\int_{-\frac{\pi}{2}}^{\frac{\pi}{2}} {\rm d} \theta\,
  p_{\rm F} \cos \theta_0 \ 
\nonumber\\
& & \qquad \times \tilde{P} ({\bf Y} ,{\bf Y}_0; t)
  \left[ \cdots \right]_{{\bf Y}_0}, \ \ 
  \label{Eq:sum-rule}
 \end{eqnarray}
where we define ${\tilde P}({\bf Y},{\bf Y}_0;t)\de y\de \theta \de t$ as the
classical probability for a particle to go from an initial position and
momentum angle of ${\bf Y}_0\equiv(y_0,\theta_0)$ on lead L to within $(\de
y,\de \theta)$ of ${\bf Y}=(y,\theta)$ on lead R in a time within $\de t$ of
$t$. For any given dot at a given energy, ${\tilde P}({\bf Y},{\bf Y}_0;t)$
has a Dirac $\delta$-function on each classical path. However, the average of
${\tilde P}$ over an ensemble of dots or over energy gives a smooth function.
We assume the classical dynamics exhibit hyperbolic chaos, thus they will be mixing.
If the dynamics are mixing on a timescale $\ll \tau_{\rm D}$, this function
will rapidly decay to a uniform distribution over phase space, so that we can
assume $ \left \langle \tilde{P} ({\bf Y};{\bf Y}_0; t) \right\rangle =
  e^{-t/\tau_{\rm D} } \cos \theta /(2 \left( W_{\rm L} + W_{\rm R} \right) \tau_{\rm D})$.
The sum of these diagonal contributions gives the Drude
conductance (also called the classical conductance)
\begin{eqnarray}
\langle g \rangle_{\rm D} = {N_{\rm L}N_{\rm R} \over (N_{\rm L} +
N_{\rm R})}. \label{eq:drude}
\end{eqnarray}

\subsection{Weak localization}
\label{sect:wl}

Systems that exhibit time-reversal symmetry (typically systems in
zero magnetic field) have correlations between paths that are the
time-reverse of each other.  If path $\gamma'$ is the time-reverse
of path $\gamma$ then $S_\gamma=S_{\gamma'}$ for all energies and
dot-shapes, and as such one could expect contributions to the
average conductance coming from these pairs of paths.  However,
looking at Eq.~\ref{Eq:g-doublesum}, we see that paths that are
the {\it exact} time-reverse of each other contribute to
reflection but not transmission, and further require that $y=y_0$
which reduces their contribution to reflection to zero. In
contrast, paths that are the {\it approximate} time-reverse of
each other, or have segments that are the approximate time-reverse
of each other, give a finite contribution to both reflection and
transmission. They are the weak localization and coherent
backscattering contributions discussed in
Ref.~\cite{Richter-Sieber}, with the latter being more carefully
analyzed in Refs.~\cite{Jac06,Rahav-Brouwer-backscatter}. Such
contributions all involve a path $\gamma$ having a loop such that
there is an {\it encounter} where the path comes very close to
itself while {\it also} having the opposite momentum (resulting in
two almost-parallel segments which are traversed in opposite
direction). Path $\gamma'$ closely follows $\gamma$, but traverses
the loop in opposite direction, which results in a crossing at the
encounter. 
For classical dynamics which exhibit hyperbolic chaos, $|S_\gamma-S_{\gamma'}| <\hbar$ 
for all energies and dot-shapes if at the encounter the two parallel segments of path
$\gamma$ are separated by a distance less than $(\lambda_{\rm
F}L)^{1/2}$. Thus such contributions survive averaging, and
summing them gives the weak localization correction to the average
conductance,
\begin{eqnarray}
\langle \de g \rangle_{\rm wl} = -{N_{\rm L}N_{\rm R} \over
(N_{\rm L} + N_{\rm R})^2} \exp[-\tEc/\tD] \times
Z(B,\gamma_\phi). \label{Eq:gwl}
\end{eqnarray}
The RMT regime is defined by $\tEc\ll \tD$ so that the factor
$\exp[-\tEc/\tD]$ equates to unity. In the classical regime ($\tEc
> \tD$), on the other hand,  the weak localization
correction is exponentially suppressed \cite{Aleiner-Larkin}. The
factor
\begin{eqnarray}
Z(B,\gamma_\phi)
&=&
 \frac{\exp[-\gamma_\phi\tilde{\tau}]}
{1+(B/B_{\rm c})^2+\gamma_\phi\tau_{\rm D}}
\label{Eq:Zwl}
\end{eqnarray}
describes the suppression due to a magnetic field and finite
dephasing. The magnetic-field induced suppression occurs with the
crossover from dynamics with time-reversal symmetry to dynamics
without time-reversal symmetry. The crossover scale $B_{\rm c}$ is
given in Eq.\ (\ref{eq:bc}).

The time scale $\tilde{\tau}$ depends on the origin of the
dephasing \cite{Bro06-quasi-ucf,pjw-decoh,wjp-decoh}. For
electron-electron interactions $\tilde\tau$ in the dot
\cite{Bro06-quasi-ucf} or dephasing leads \cite{pjw-decoh} it is
of order the Ehrenfest time, while for dephasing due to microwave
noise or charge-fluctuations on gates near the dot it is given by
the logarithm of the ratio of the noise correlation length, $\xi$,
to the dot size. See Ref.~\cite{wjp-decoh} for an extensive
discussion of the nature of $\xi$ and $\tilde \tau$ for all these
sources of noise.

\subsection{Universal conductance fluctuations (UCFs)}

The conductance of an individual quantum dot has a characteristic
dependence on energy, applied magnetic field, and deformations of
the dot's shape (which can be controlled by the voltages applied
to the side- or top gates providing the electrostatic
confinement). The magnitude of these fluctuations can be
characterized by the variance of the  conductance, which
``universally'' has a magnitude of order $(e^2/h)^2$ (meaning the
variance of the dimensionless conductance is of order 1). In the
absence of dephasing, semiclassics predicts UCFs in a chaotic
quantum dot with no spatial symmetries given by
\cite{Brouwer-Rahav-ucfs,Brouwer-Rahav-ucfs2}
\begin{eqnarray}
{\rm var}(g) = {N_{\rm L}^2 N_{\rm R}^2 \over (N_{\rm L}+N_{\rm
R})^4} \times Z_{\rm ucf}(B),
\end{eqnarray}
where
\begin{eqnarray}
Z_{\rm ucf}(B)=
1 +
{\e^{-\tEo/\tau_{\rm D}}
\over [1+(B/B_{\rm c})^2]^2}
+ {1-\e^{-\tEo/\tau_{\rm D}}
\over 1+(B/B_{\rm c})^2}
\label{Eq:UCF-suppression}
\end{eqnarray}
describes the effect a perpendicular magnetic field $B$. Thus for
$B \ll B_{\rm c}$ or $B \gg B_{\rm c}$ the magnitude of the UCFs
is independent of the Ehrenfest time $\tEo$. However, the form of
the crossover between the two limiting cases depends on the
Ehrenfest time; the crossover is Lorentzian in the classical limit
($\tEo \gg \tau_{\rm D}$) and Lorentzian-squared in the RMT limit
($\tEo \ll \tau_{\rm D}$). The effect of dephasing on UCFs is more
complicated \cite{Bro06-quasi-ucf}  and will not be considered
here.

\subsection{Adding spatial symmetries to the semiclassical method}

We start our discussion of spatial symmetries by explaining the
sketches in Figs.~\ref{Fig:leftright}-\ref{Fig:updown}. The
relationship between the depicted paths and real classical paths
in a chaotic dot is illustrated in Fig.~\ref{Fig:example}. The
abstract sketches in Figs.~\ref{Fig:leftright}-\ref{Fig:updown}
neglect all details of the paths except the following essential
ones:
\begin{itemize}
\item We show that a path crosses the symmetry axis only when it
is necessary to understand the nature of the path. For example the
paths in Fig.~\ref{Fig:leftright}(a) going from  lead L to lead R
are shown to cross the symmetry axis once, when they may in fact
cross the symmetry axis any (odd) number of times. Indeed a
typical path will cross the symmetry axis every few bounces, and
will bounce very many times before escaping into a lead.

\item We show when a path comes extremely close in phase space to its
    symmetric partners (related to it by spatial symmetry).  These {\it
    effective encounters} play a similar role to those discussed above
    for weak localization.
\end{itemize}
In the figures, the right-hand sketches show all paths folded (via
the symmetry) into half the system, which emphasizes when a path
comes close to its spatially symmetric partner (i.e., its mirror
image in systems with a mirror symmetry). As is the case with
time-reversal symmetry, pairs of paths that are exactly symmetric
have a negligible contribution to the conductance. However, pairs
of {\it approximately} symmetric paths [such as shown in
Fig.~\ref{Fig:leftright}(a)] sum to give a finite contribution.
This is also the case for pairs of paths which have some segments
which are approximately the same and other segments which are
approximately symmetric to each other [such as shown in
Fig.~\ref{Fig:leftright}(b)].

\section{Left-right mirror symmetry}
\label{sect:LR}

\begin{figure}
\centerline{\hbox{\includegraphics[width=1.0\columnwidth]{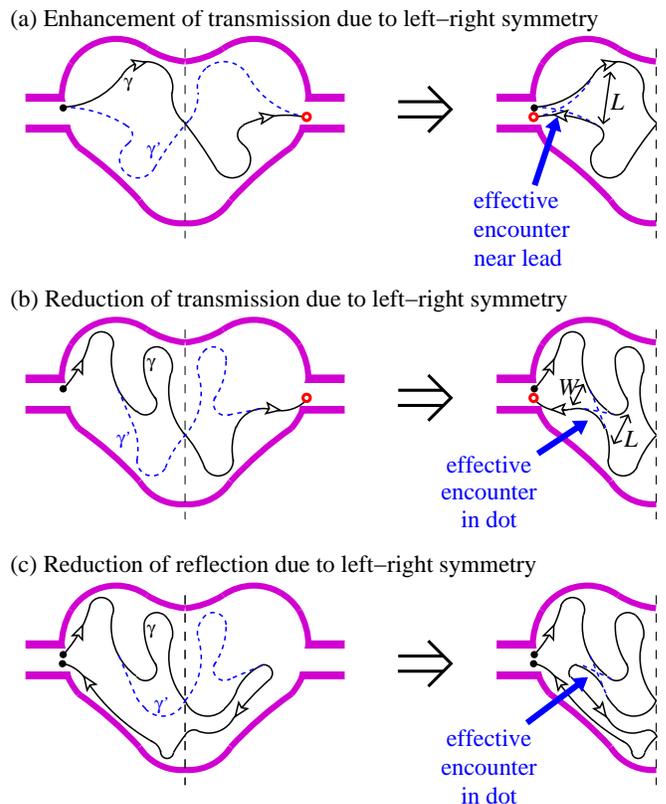}}}
\caption[]{\label{Fig:leftright} 
(colour online). Contributions to weak
localization due to a left-right mirror symmetry. When we fold the
paths into one half of the dot (as shown in the right-hand
sketches), paths hitting the symmetry axis are specularly
reflected; however, we must keep track of whether the path escapes
into the right lead. Thus in those sketches, paths that end on the
left lead are marked with solid circles, while those that end on
the right lead are marked with open (red) circles. The sketch in
(a) is an enhanced forward-scattering contribution, because it
enhances transmission through the dot. }
\end{figure}

First we analyze the case of a perfectly left-right symmetric dot
with symmetric leads, $W_{\rm L}=W_{\rm R}$  and therefore $N_{\rm
L}=N_{\rm R} \equiv N$. There are three interference contributions
to transmission and reflection that exist only because of the
symmetry. Two of these contribute to transmission
[Fig.~\ref{Fig:leftright}(a),(b)], while the third contributes to
reflection [Fig.~\ref{Fig:leftright}(c)].

\subsection{Enhanced forward scattering}
 \label{sect:leftright-contribs-a}

The contribution of paths of the type in
Fig.~\ref{Fig:leftright}(a) have an effective encounter which is
close to the leads. This is analogous to the
coherent-backscattering contribution in Ref.~\cite{Jac06} (or the
$D_2,D_3$-contributions to the shot noise in Ref.~\cite{wj-fano})
and we therefore use the same method to treat such paths here.
First we note that the behaviour of path $\gamma'$ is completely
determined by that of path $\gamma$; the two paths have the same
amplitudes, $A_{\gamma'}=A_\gamma$ and the action difference
between them is $S_{\gamma}-S_{\gamma'} = (p_{0\perp}+m\Lambda
r_{0\perp})r_{0\perp}$, where $(r_{0\perp},p_{0\perp})$ is the
component of ${\bf Y}-{\bf Y}_0$ which is perpendicular to the
direction of path $\gamma$ at ${\bf Y}$. This action difference 
requires hyperbolic chaos and is 
discussed in detail in the context of coherent backscattering in
Ref.~\cite{Jac06}. Using the sum rule in Eq.~(\ref{Eq:sum-rule}),
we see that the contribution of such paths to the conductance is
\begin{eqnarray}
\label{eq:g_LRa}
\langle \de g \rangle_{\rm LR:a}
&=& (2\pi \hbar)^{-2}
\!\int_{\rm L} \!\! \rmd {\bf Y}_0 \!\int_{\rm R} \!\!  \rmd {\bf Y}
\int_0^\infty \!\! \rmd t
\\
& & \quad \times p_{\rm F} \cos \theta_0 \,\langle \tilde{P}({\bf Y},{\bf
Y}_0;t)\rangle \; {\rm Re}\big[\e^{\rmi (S_{\gamma}-S_{\gamma'})
/\hbar}\big] . 
\nonumber 
\end{eqnarray}

To perform the average we proceed as for coherent backscattering
\cite{Jac06}. Starting at the leads, we define $T'_{W}(r_{0\perp},
p_{0\perp})$ and $T'_{L}(r_{0\perp}, p_{0\perp})$ as the time at
which the perpendicular distance between $\gamma$ and $\gamma'$
reaches $W$ and $L$, respectively. For times less than
$T'_{W}(r_{0\perp}, p_{0\perp})$, the path segments are paired,
and thus their joint survival probability is the same as for a
single path (if one path does not touch a lead then the other also
does not). For times longer than this the path segments escape
independently. We only have contributions if paths $\gamma$ and
$\gamma'$ diverge from each other in phase space before
reconverging at the R lead. For the reconvergence to occur the
paths must diverge to a distance of order $L$ (on shorter scales
the dynamics are hyperbolic so the reconvergence of diverging
paths is not allowed). This means that the $t$-integral in
Eq.~(\ref{eq:g_LRa}) must have a lower cut-off at
$2T'_L(r_{0\perp}, p_{0\perp})$.

Now we write the ${\bf Y}_0$ integral  in Eq.\ (\ref{eq:g_LRa})
 in terms of the relative coordinates $(r_{0\perp}, p_{0\perp})$,
using $p_{\rm F} \cos\theta_0 \rmd {\bf Y}_0 = \rmd y_0\rmd
(p_{\rm F} \sin\theta_0 ) = \rmd r_{0\perp} \rmd p_{0\perp}$. Then
the ${\bf Y}$ integral becomes explicitly independent of
$(r_{0\perp},p_{0\perp})$ and we can write
\begin{eqnarray}
& & \int_{\rm R} \! \rmd {\bf Y}
\int_{2T'_L}^\infty \rmd t
\langle \tilde{P}({\bf Y},{\bf Y}_0;t)\rangle
\nonumber \\
& & \quad =
   {1\over 2}
\exp[-T'_W/\tD -2(T'_L-T'_W)/\tD] ,\
\, \qquad
\label{Eq:int-of-P}
\end{eqnarray}
where $T'_{L,W}$ are shorthand for $T'_{L,W}(r_{0\perp}, p_{0\perp})$
and we used the fact that $N_{\rm L}=N_{\rm R}$ to get the prefactor of a half 
(more generally the prefactor would be $N_{\rm R}/(N_{\rm L}+N_{\rm R})$).
For small $(p_{0\perp}+ m\Lambda r_{0\perp})$ we get
\begin{eqnarray}
T'_L(r_{0\perp}, p_{0\perp}) &\simeq& \Lambda^{-1} \ln \left[ {
m\Lambda L \over |p_{0\perp}+ m\Lambda r_{0\perp}| }\right] ,
\qquad \label{eq:T'_L}
\end{eqnarray}
with $T'_W(r_{0\perp}, p_{0\perp})$ given by the same formula
where $L$ is replaced by $W$. Thus we see that
$2(T'_L-T'_W)=\tEc-\tEo$, which generates a factor
$\exp[-(\tEc-\tEo)/\tD]$ and is independent of
$(r_{0\perp},p_{0\perp})$. Substituting these results into Eq.\
(\ref{eq:g_LRa})  for $\langle \de g\rangle_{\rm LR:a}$, and
recalling that $p_{\rm F} \cos\theta_0 \rmd {\bf Y}_0 = \rmd
y_0\rmd (p_{\rm F} \sin\theta_0 ) = \rmd r_{0\perp} \rmd
p_{0\perp}$, we get 
\begin{eqnarray}
\langle \de g \rangle_{\rm LR:a}
&=& {\e^{-(\tEc-\tEo)/\tD} \over 2(2\pi \hbar)^2}
\!\int_{\rm L} \!\! \rmd r_{0\perp} \rmd
p_{0\perp} \, \e^{-T'_W/\tD}
\nonumber \\
& & \quad \times {\rm Re}\big[\exp[ \rmi (p_{0\perp}+m\Lambda
r_{0\perp})r_{0\perp}/\hbar]\big] . \quad
\end{eqnarray}
We next make the substitution
$\tilde{p}_0=p_{0\perp}+m\Lambda r_{0\perp}$ and evaluate the
$r_{0\perp}-$integral over a range of order $W$. We take the limits
on the resulting $\tilde{p}_0$-integral to $\pm \infty$.
One can use Euler $\gamma$-functions to show that
\begin{eqnarray}
& & \hskip -10mm 
\int_{-\infty}^\infty \rmd \tilde{p}_0
{2\hbar \sin(\tilde{p}_0 W/\hbar)\over \tilde{p}_0}
 \left| \tilde{p}_0 \over m\Lambda W \right|^{1/(\Lambda\tD)} 
\nonumber \\
&=& 2\pi \hbar \left(\hbar \over m\Lambda W^2 \right)^{1/(\Lambda\tD)},
\label{eq:tilde_p-int1}
 \end{eqnarray}
the trick in evaluating this integral is to write it as the sum of two integrals from 
0 to $\infty$.
The factor of $[\hbar/(m\Lambda
W^2)]^{1/(\Lambda\tD)} \simeq \exp[-\tEo/\tD]$, where we have
dropped ${\cal O}(1)$-terms inside a logarithm. This cancels the
$\tEo$-term in the factor $\exp[-(\tEc-\tEo)/\tD]$. Collecting
terms and evaluating some trivial integrals leads to 
\begin{eqnarray}
\langle \de
g \rangle_{\rm LR:a} = {1 \over 2} \exp[-\tEc/\tD].
\end{eqnarray}

 \subsection{Uniform reduction of transmission and reflection}
 \label{sect:leftright-contribs-bc}

We now turn to contributions with effective encounters in the dot,
Fig.~\ref{Fig:leftright}(b),(c). We define an effective encounter as being
``in the dot''  if the parts of path $\gamma$ on either side of the encounter
separate to a distance of at least $W$ before touching any leads. In this
case, the ends of path $\gamma$ are independent of each other, with no
correlations in their position, momentum or the time at which they hit a
lead. As in the case of the enhanced forward scattering, the path $\gamma$
and $\gamma'$ must in fact separate to a distance of order $L$ before they
can reconverge. Thus we have two minimal times; $T_{\rm L}(\epsilon)$ for
$\gamma$ and $\gamma'$ to separate before reconverging [to the right of the
encounter in Fig.~\ref{Fig:leftright}(b)], and $T_{\rm W}(\epsilon)$ for the
two parts of $\gamma$ to separate enough to escape independently into the
leads [to the left of the encounter in Fig.~\ref{Fig:leftright}(b)].

Assuming the encounter is close (i.e.~path $\gamma'$ crosses
itself with very small $\epsilon$) and that the classical dynamics
are hyperbolic at such small scales, one estimates
   \begin{eqnarray}\label{TW}
 T_W(\epsilon) &  \simeq  & \Lambda^{-1} \ln [\epsilon^{-2}(W/L)^2],
 \end{eqnarray}
with $T_L(\epsilon) \simeq  \Lambda^{-1} \ln [\epsilon^{-2} ]$.
The action difference is $S_\gamma-S_{\gamma'} = E_{\rm F}
\epsilon^2/\Lambda$, just as  for weak localization
\cite{Sie01,Richter-Sieber}. To evaluate the contributions of such
paths, the probability ${\tilde P} ({\bf Y},{\bf Y}_0;t)$ in the
sum rule of Eq.~(\ref{Eq:sum-rule}) is restricted to paths which
cross themselves, so that we can write
\begin{eqnarray}
 {\tilde P} ({\bf Y},{\bf Y}_0;t) &=&
 \int_{\rm ps} {\rm d} {\bf R}_2 {\rm d} {\bf R}_1
 \tilde{P} ({\bf Y},{\bf R}_2;t-t_2) 
\nonumber \\
& & \times
 \tilde{P}({\bf R}_2,{\bf R}_1;t_2-t_1)
 \tilde{P}({\bf R}_1,{\bf Y}_0;t_1) \,, \qquad
\end{eqnarray}
where ${\rm ps}$ indicates that the integrals are over the phase space of the
dot. Here, we use ${\bf R}=({\bf r},\phi)$, $\phi \in [-\pi,\pi]$ for
phase-space points inside the dot, while ${\bf Y}$ lies on the lead as
before. We then restrict the probabilities inside the integral to
trajectories which cross their mirror image at phase-space positions ${\bf
R}_{1,2}$ with the first (second) visit to the crossing occurring at time
$t_1$ ($t_2$).  We can write ${\rm d} {\bf R}_2 = v_{\rm F}^2 \sin \epsilon
{\rm d}t_1 {\rm d}t_2{\rm d} \epsilon$ and set ${\bf R}_2 =({\bf
r}'_1,\phi'_1\pm \epsilon)$, where ${\bf r}'_1$ and $\phi'_1$ are the mirror
images of position ${\bf r}_1$ and angle $\phi_1$ (thus
$\phi'_1=\pi-\phi_1$). Then the correction to the dimensionless conductance
is $ \langle \de g \rangle_{\rm LR:b} = (\pi \hbar)^{-1} \int_{\rm L} \! {\rm
d} {\bf Y}_{0}\int {\rm d} \epsilon \,
 {\rm Re}\big[e^{i(S_{\gamma}-S_{\gamma'}) /\hbar }\big]
 \big\langle  F ( {\bf Y}_{0}, \epsilon)  \big\rangle$,
with
 \begin{eqnarray}
 F( {\bf Y}_{0}, \epsilon)  &=&
 2v_{\rm F}^2  \sin \epsilon
  \int_{T_L+T_W} ^{\infty} \!  {\rm d } t
 \int_{T_ L+\frac{T_W}{2}}^{t-\frac{T_W }{2} } \!   {\rm d} t_2
 \int_{\frac{T_W}{2}}^{t_2-T_L}  \!   {\rm d} t_1
\nonumber \\
& &\times
 p_{\rm F}\cos \theta_0 \int_{\rm R} \!  {\rm d} {\bf Y}
 \int_{\rm ps} \! {\rm d} {\bf R}_1
 \tilde{P}({\bf Y} ,{\bf R}_2 ;t-t_2)
\nonumber \\
& &\times
 \tilde{P}({\bf R}_2,{\bf R}_1;t_2-t_1)
 {\tilde P}({\bf R}_1,{\bf Y}_0;t_1) .
\label{Eq:F}
 \end{eqnarray}
Comparison with Eq.~(34) of Ref.~\cite{Jac06} shows that this has the same
form as the weak localization correction. Thus we only briefly summarize the
remainder of the derivation here. Assuming phase space ergodicity  for the
system, we have
\begin{eqnarray}
\left\langle  F( {\bf Y}_{0}, \epsilon)  \right\rangle
&=&  {2v_{\rm F}^2\tD^2 \over 2\pi A}
{N_{\rm R} \over  N_{\rm L}+N_{\rm R}}
\nonumber \\
& &\times
p_{\rm F} \cos\theta_0
\,\sin \eps \,\exp \big[-T_L(\eps)/\tau_{\rm D}\big], \qquad
\label{Eq:F-result}
\end{eqnarray}
with $A$ being the real space volume occupied by the system (the area of the dot).
Then
the integral over $\eps$ in $\langle \de g \rangle_{\rm LR:b}$
takes the form
${\rm Re} \int_0^\infty \rmd \eps \,
 \eps^{1+ 2/(\Lambda \tau_{\rm D})}
 \,\exp[\rmi E_{\rm F} \eps^2/(\Lambda \hbar)]$,
where we have assumed $\eps \ll 1$ \cite{Jac06}.
 The substitution $z=E_{\rm F} \eps^2/(\Lambda \hbar)$ immediately
 yields a dimensionless integral
 and an exponential term, $\e^{-\tau_{\rm E}^{\rm cl}/\tau_{\rm D}}$
 (neglecting as usual ${\cal O}[1]$-terms in the logarithm in
 $\tau_{\rm E}^{\rm cl}$).
 From this analysis, we find that this interference correction
to the conductance
 is given by
$\langle \de g \rangle_{\rm LR:b} = -{1 \over 4} \exp[-\tEc/\tD]$.

The backscattering contribution in Fig.~\ref{Fig:leftright}(c)
does not directly enter the conductance.  However, we must
evaluate it if we wish to explicitly show that the theory
conserves particles. If interference causes an enhancement of
transmission then there must be an associated reduction of
reflection; it is the paths in Fig.~\ref{Fig:leftright}(c) that
give this reduction of reflection, which we denote by $\langle \de
R \rangle_{\rm LR:c}$. Inspecting the figure we see that this
contribution is of the same form as Eq.~(\ref{Eq:F}), but with the
${\bf Y}$-integral now over taken over lead L instead of lead R.
Thus we find that $\langle \de R \rangle_{\rm LR:c} = \langle \de
g \rangle_{\rm LR:b}$ so $\langle \de g \rangle_{\rm LR:a}
+\langle \de g \rangle_{\rm LR:b} +\langle \de R \rangle_{\rm
LR:c} =  0$, which confirms that the theory conserves particles.

\subsection{Asymmetry and dephasing in the dot}

\begin{figure*}
\centerline{\hbox{\includegraphics[width=0.7\textwidth]{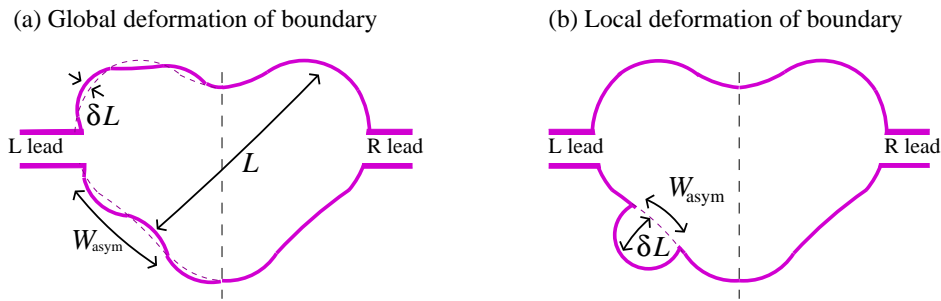}}}
\caption[]{\label{Fig:asymm} 
(colour online).  Two ways in which the left half of a
quantum dot can be deformed away from symmetry with the right
half. In (a) the left half is deformed over a long range $W_{\rm
asym} \gg \lambda_{\rm F}$
 by a
gentle ripple of small magnitude $\de L < \lambda_{\rm F}$. A
special case of this is where the whole left-half of the dot is
deformed outwards by a distance of order $\de L$, then $W_{\rm
asym}\sim L$. In (b) a small part of the boundary of the left half
(of width $W_{\rm asym}$) is deformed by a distance $\de L \gg
\lambda_{\rm F}$. In the former case the parameter which controls
the deviation from symmetry is $\de L/\lambda_{\rm F}$ while in
the latter case it is $W_{\rm asym}/L$. }
\end{figure*}

We now turn to the effect of deformations of the dot's shape which
break the left-right mirror symmetry. We consider two cases; the
first is a small deformation of a large proportion of the dot's
boundary, see Fig.~\ref{Fig:asymm}(a), and the second is a large
deformation of a small proportion of the dot's boundary,  see
Fig.~\ref{Fig:asymm}(b). Other types of deformation can be
understood by applying the methodology we introduce for these two
cases. In general, the effect of asymmetries on symmetry-induced
interference corrections to the conductance is similar to the
effect of magnetic fields and dephasing on weak localization for a
completely asymmetric dot \cite{Richter-Sieber}.

Consider the left side of the dot being deformed such that the
boundary is rippled, see the sketch in Fig.~\ref{Fig:asymm}(a).
The ripple does not change in time, and has a small amplitude $\de
L < \lambda_{\rm F}$ and long range $W_{\rm asym} \gg \lambda_{\rm
F}$ (the latter condition ensures that there is no problem using
the semiclassical theory). The simplest such case is when the
whole left side of the dot deformed slightly outwards, which
corresponds to $W_{\rm asym} \sim L$.  Then a path in the left
half of the dot which bounces from this portion of the dot
boundary will travel a distance of order $\de L$ further than its
partner in the right half.  We assume that this small difference
has no significant effect on the classical trajectories but
changes the phase acquired along the path and thereby modifies the
interference with its symmetry-related partner. Paths are chaotic
and regularly switch between the left and right halves of the dot.
Thus it is reasonable to assume the length difference between a
path segment of time $t \gg \tau_0$ and its mirror image is the
sum of many independent random lengths distributed such that
$\langle \de L\rangle =0$ and $p_{\rm F}^2\langle\de L^2\rangle
\ll \hbar^2$, where the average is over the whole circumference of
the dot. Thus the action difference for a path segment of time $t$
(which therefore bounces $t/\tau_0$ times) and its mirror image is
$\big\langle \exp[\rmi (S_\gamma-S_{\gamma'})/\hbar]\big\rangle =
\prod_{n=1}^{t/\tau_0} \langle \exp[\rmi p_{\rm F} \de L/\hbar]
\rangle_n$. Irrespective of the probability distribution of each $\de L$,
we can use central limit theorem to tell us that
the probability distribution for $ (S_\gamma-S_{\gamma'})$ is a Gaussian
(Bell-curve) with variance given by the sum of the variances of the individual
$\de L$s; thus for times $t \gg \tau_0$ we have
\begin{eqnarray}
{\rm var} [S_\gamma-S_{\gamma'}] \simeq (t/\tau_0) \langle \de L
\rangle/\lambda_{\rm F}^2.
\end{eqnarray}
Thus $\big\langle \exp[\rmi (S_\gamma-S_{\gamma'})/\hbar]\big\rangle$,
which is given by a Gaussian integral over $S_\gamma-S_{\gamma'}$.
Evaluating this integral, we find that 
\begin{eqnarray}
\big\langle \exp[\rmi (S_\gamma-S_{\gamma'})/\hbar]\big\rangle
= \exp[-\gamma_{\rm asym}t]
\label{Eq:exp-suppression}
\end{eqnarray}
with decay rate
\begin{eqnarray}
\label{Eq:gamma-asym-gentle-ripple}
\gamma_{\rm asym}= {\langle \de L^2\rangle/(2\tau_0\lambda_{\rm F}^2)}.
\end{eqnarray}
Now we turn to the question of when such random phase differences
occur for the interference contributions to conductance. In the
contributions we discuss, paths $\gamma$ and $\gamma'$ are nearly
identical when they enter the effective encounter, while $\gamma$
is paired with the image of $\gamma'$ when they exit the
encounter.  The acquisition of the random phases discussed above
only starts to occur once path $\gamma$ and $\gamma'$ are a
distance $W_{\rm asym}$ or more apart. When the two paths are
closer than $W_{\rm asym}$, they do acquire extra phases due to
the deformation; however, the phase {\it difference} between them
is negligible. We assume that $W_{\rm asym} \gg \lambda_{\rm F}$,
so we can avoid worrying about diffractive scattering from the
deformed boundary (it remains smooth at the scale of the Fermi
wavelength).

We first consider the effect of asymmetry on the coherent
forward-scattering contribution of Fig.\ \ref{Fig:leftright}(a),
discussed in Sect.~\ref{sect:leftright-contribs-a}. In this
contribution, almost everywhere the paths $\gamma$ and $\gamma'$
are the mirror image of each other; the exception is an interval
of duration  $(t-2T'_{W_{\rm asym}})$ where the  paths separate to a distance
of order $W_{\rm asym}$ [here $T'_{W_{\rm asym}}$ is given in Eq.~(\ref{eq:T'_L}) with $W_{\rm asym}$ in place of $L$]. Thus
the integral over time $t$ in Eq.~(\ref{Eq:int-of-P}) acquires
another factor of the form $\exp \big[ -\gamma_{\rm
asym}(t-2T'_{W_{\rm asym}})\big]$. Thus asymmetry has a similar effect as
dephasing, which contributes a factor of the form
$\exp[-\gamma_\phi(t-2T'_\xi)]$ \cite{pjw-decoh,wjp-decoh}; here
$T'_\xi$ is given by Eq.~(\ref{eq:T'_L}), with the
noise correlation length $\xi$ in place of $L$ (see
Section~\ref{sect:wl} and references therein). The integral in
Eq.~(\ref{Eq:int-of-P}) is therefore replaced by
\begin{eqnarray}
& & \hskip -8mm \int_{\rm R} \! \rmd {\bf Y}
\int_{2T'_L}^\infty \rmd t
\langle {\tilde P}({\bf Y},{\bf Y}_0;t)\rangle
\nonumber \\
& & \qquad \times \exp[-(\gamma_{\rm asym}+\gamma_\phi)t]
\exp[\gamma_{\rm asym}T'_{L}+\gamma_\phi T'_{\xi}]
\nonumber \\
&=& {1\over 2} \exp[-T'_W/\tD -2(T'_L-T'_W)/\tD] \, 
\nonumber \\
& & \qquad \times 
{\exp[
-\gamma_\phi{\tilde \tau} - \gamma_{\rm asym}{\tilde \tau}_{\rm
asym} ] \over 1+(\gamma_{\rm asym}+\gamma_\phi)\tD},
\label{Eq:int-of-P-with-suppression}
\end{eqnarray}
where for compactness we define two effective phase correlation
times $\tilde\tau = 2(T'_L-T'_{\xi})$ and $\tilde\tau_{\rm asym} =
2(T'_L-T'_{W_{\rm asym}})$ for dephasing and asymmetries,
respectively, which take the explicit form
\begin{eqnarray}
\tilde\tau &=& \Lambda^{-1} \ln \big[ (L/\xi)^2 \big],
\nonumber \\
\tilde\tau_{\rm asym} &=& \Lambda^{-1} \ln \big[ (L/W_{\rm asym})^2 \big].
\label{Eq:tilde-taus}
\end{eqnarray}
If the length scale of the deformation of the boundary, $W_{\rm
asym}$, is of order $L$ then we can treat $\tilde\tau_{\rm asym}
\simeq 0$. Similarly, for dephasing, we can treat $\tilde\tau
\simeq 0$ for $\xi \sim 0$; however, for electron-electron
interactions in the dot, $\tilde\tau$ is given by the Ehrenfest
time $\tEc$ (see
 Ref.~\cite{Bro06-quasi-ucf}
and Sect. IIIA of Ref.~\cite{wjp-decoh}, which reviews that work
using a similar notation as the present article).

Compared to Eq.~(\ref{Eq:int-of-P}),
Eq.~(\ref{Eq:int-of-P-with-suppression}) contains an additional
constant factor. The remaining derivation of the contribution of
coherent forward scattering can therefore proceed exactly as in
Section~\ref{sect:leftright-contribs-a}. The final result is
$\langle \de g \rangle_{\rm LR:a} = {1\over 2} \exp[-\tEc/\tD]
\times Z_{\rm LR}(\gamma_{\rm asym},\gamma_\phi)$, with
\begin{eqnarray}
Z_{\rm LR}(\gamma_{\rm asym},\gamma_\phi) = {\exp[
-\gamma_\phi\tilde{\tau}-\gamma_{\rm asym}\tilde{\tau}_{\rm asym}] \over 1+(\gamma_{\rm
asym}+\gamma_\phi)\tD}. 
\label{Eq:Z_LR1}
\end{eqnarray}

Turning to the other interference contributions for a left-right
symmetric dot, Fig.~\ref{Fig:leftright}(b),(c), we carry out a
similar analysis for the parts of the paths which are approximate
mirror images of each other.  Measuring from the encounter, the
paths diverge to a distance $W_{\rm asym}$ apart in a time
$T_{W_{\rm asym}}/2$, at which point the asymmetry-induced
suppression sets in. Similarly, the paths diverge to a distance
$\xi$ apart in a time $T_\xi/2$, at which point the
dephasing-induced suppression sets; here the time $T_\xi$ and
$T_{\rm W_{\rm asym}}$ are  given by Eq.~(\ref{TW}) with $\xi$ and
$W_{\rm asym}$, respectively, in place of $W$. The asymmetry
suppresses the interference term by a factor $\exp [-\gamma_{\rm
asym}(t_2-t_1-T_{W_{\rm asym}})]$, while dephasing results in a
suppression of the form $\exp [-\gamma_{\phi}(t_2-t_1-T_{\xi})]$.
Repeating the analysis in Section~\ref{sect:leftright-contribs-bc}
with these extra terms, we find that $\langle \de g \rangle_{\rm
LR:b}$ and $\langle \de R \rangle_{\rm LR:c}$ are also suppressed
by factors of $Z_{\rm LR}(\gamma_{\rm asym},\gamma_\phi)$.

It is noteworthy that the expression for $\langle \de g \rangle_{\rm LR:a}$,
$\langle \de g \rangle_{\rm LR:b}$, and $\langle \de R \rangle_{\rm LR:c}$
all are  independent of the magnetic field, which is quite unlike the
weak localization correction given by Eqs.~(\ref{Eq:gwl},\ref{Eq:Zwl}). The
reason for the magnetic-field independence lies in the orientation of
trajectory segments in symmetry-related pairs of paths $\gamma$, $\gamma'$.
In all contributions, when both paths are not approximately the same, they
are approximately related by taking the mirror image and time-reverse. While
the flux enclosed by a path changes its sign by mirror reflection or time
reversal, the combined action of both operations results in a path which
encloses the same flux as the original path. As a result, the phase induced
by the  magnetic field is exactly the same for path $\gamma$ and $\gamma'$
and drops out of the interference contributions, since the latter only
contain the phase difference. This is in stark contrast to the
weak localization correction, where the symmetric partner of a path is just
its time-reverse, which means the flux enclosed by $\gamma$ and $\gamma'$ is
opposite (in section~\ref{sect:inversion} we encounter a similar situation in
a dot with inversion symmetry).

We now turn to local deformations as sketched in
Fig.~\ref{Fig:asymm}(b).  If neither path $\gamma$ nor $\gamma'$
go into the deformed region while they are approximate mirror
images of each other, then there is no phase shift, and the
contribution is the same as for a perfectly symmetric system.
However, if one of the paths enters the deformed region while
paired with the mirror image of the other it will immediately
acquire a phase difference of $(S_\gamma-S_{\gamma'})/\hbar \gg
1$, and will fluctuate strongly under energy or ensemble
averaging. The resulting average $\big\langle \exp[\rmi
(S_\gamma-S_{\gamma'})/\hbar]\big\rangle$ is still given by
Eq.~(\ref{Eq:exp-suppression}), but the decay rate now takes the
form
\begin{eqnarray}
\gamma_{\rm asym}= {W_{\rm asym}/(2\tau_0C)} = \tD^{-1}\, (W_{\rm asym}/W),
\label{Eq:gamma-asym-local-deform}
\end{eqnarray}
where $C$ is the circumference of the cavity and $W_{\rm asym}$ is the width
of the deformation [see Fig.~\ref{Fig:asymm}(b)]. This asymmetry-induced
suppression can only set in once the two paths are a distance of order
$W_{\rm asym}$ apart (only then can one path hit the deformation while the
other does not), which results in an additional factor $\exp[-(T_L-T_{W_{\rm
asym}})/\tau_{\rm D}]$, where $T_{W_{\rm asym}}$ is given by Eq.~(\ref{TW})
with $W_{\rm asym}$ in place of $W$. Following through the calculation we
arrive once again at Eq.~(\ref{Eq:Z_LR1}) but now with $\gamma_{\rm asym}$
given by Eq.~(\ref{Eq:gamma-asym-local-deform}); $\tilde{\tau}$ and
$\tilde{\tau}_{\rm asym}$ are unchanged and given by
Eq.~(\ref{Eq:tilde-taus}).

Intriguingly,  unlike the dephasing rate
Eq.~(\ref{Eq:gamma-asym-gentle-ripple}) induced by the global
asymmetry in  Fig.~\ref{Fig:asymm}(a),
  the dephasing
rate  Eq.~(\ref{Eq:gamma-asym-local-deform}) induced by the
localized asymmetry in Fig.~\ref{Fig:asymm}(b) is
wavelength-independent. Thus $W_{\rm asym}$ can be a classical
scale (much bigger than $\lambda_{\rm F}$ while remaining less
than $W$) without destroying the symmetry-induced effects.  For
example, even if $W_{\rm asym} \sim W$ we have $\gamma_{\rm
asym}\tD \sim 1$ and the symmetry-induced effect is only
suppressed by a factor of two (then the exponential suppression,
$\exp[-\gamma_{\rm asym}\tilde{\tau}_{\rm asym}]$, is almost
irrelevant since the exponent contains a logarithm with a small
prefactor $(\Lambda \tD)^{-1} \ll 1$).

\subsection{The conductance of a left-right symmetric dot
including interference corrections}

The total interference correction to the conductance induced by
the left-right mirror symmetry is given by $\langle \de
g\rangle_{\rm LR}=\langle \de g\rangle_{\rm LR:a}+\langle \de
g\rangle_{\rm LR:b}= {1\over 4} \exp[-\tEc/\tD] \times Z_{\rm
LR}(\gamma_{\rm asym},\gamma_\phi)$, with $Z_{\rm LR}(\gamma_{\rm
asym},\gamma_\phi)$ given by Eq.~(\ref{Eq:Z_LR1}).
Thus a quantum dot with symmetric leads has
a total average conductance of
\begin{eqnarray}
\langle g\rangle_{\rm LR} &=& \langle  g\rangle_{\rm D}+
\langle \de g\rangle_{\rm wl}+\langle \de g\rangle_{\rm LR}
\nonumber \\
&=& {N \over 2} + {1 \over 4}\exp[-\tEc/\tD] \big[Z_{\rm
LR}(\gamma_{\rm asym},\gamma_\phi) - Z(B,\gamma_\phi) \big] 
\nonumber \\
& & \ +\
{\cal O}[N^{-1}]. \label{Eq:gLRsym}
\end{eqnarray}
In the absence of spatial symmetries, the higher-order
weak localization corrections (order $N^{-1}$, etc) were evaluated
in Ref.~\cite{haake-conductance}. Symmetry also induces
corrections to these orders, but they are beyond the analysis that
we carry out here.

As noted in Refs.~\cite{Baranger-Mello,Gopar96}, for perfect
left-right mirror symmetry and perfect time-reversal symmetry (no
magnetic field) the two interference corrections to the
conductance exactly cancel, so that the conductance is given by
the classical Drude conductance (\ref{eq:drude}). Interestingly,
our calculation shows that this statement remains true for
arbitrary dephasing rate and Ehrenfest time, which enter all
symmetry-induced corrections with the same functional dependence.
However, a magnetic field suppresses the negative
weak localization correction, but does not affect the corrections
from mirror symmetry. Thus the  magneto-conductance curve of a
symmetric dot looks like the shifted curve of an asymmetric dot;
the shift is positive and equals $1/4$ in the RMT limit
($\tau_\phi\gg\tD\gg \tau_E$). For $W_{\rm asym} \sim L$ (relevant
for deformations in Fig.~\ref{Fig:asymm}(a)) the correction has a
Lorentzian dependence on the magnitude of the deformation $\de L$.
For $W_{\rm asym} \ll L$, the form of the suppression is more
complicated and given in Eq.~(\ref{Eq:Z_LR1}).

\section{Inversion symmetry}
\label{sect:inversion}

\begin{figure}
\centerline{\hbox{\includegraphics[width=1.0\columnwidth]{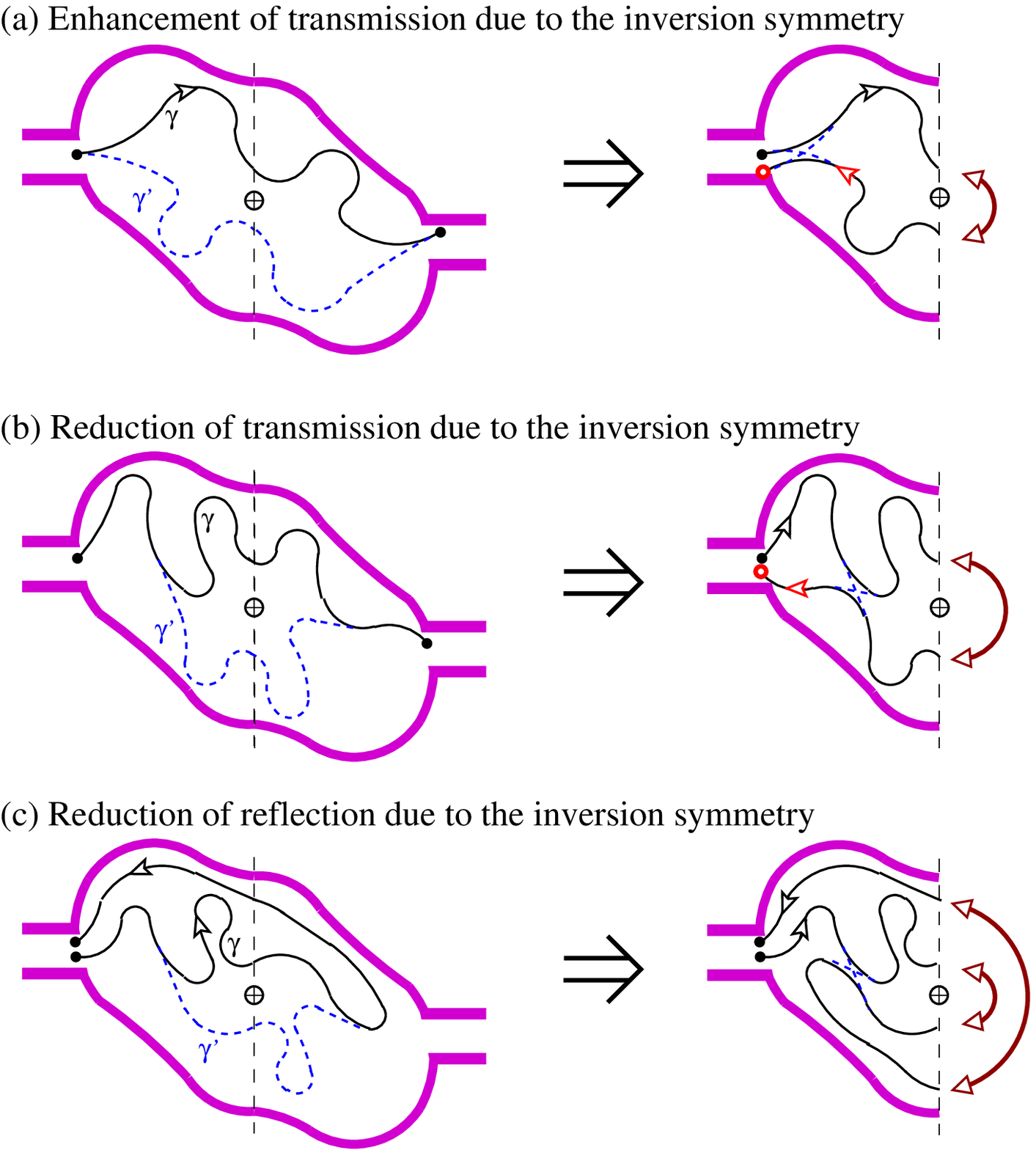}}}
\caption[]{\label{Fig:inversion} 
(colour online). Contributions to weak
localization due to an inversion symmetry, which is equivalent to
symmetry under a $180^\circ$ rotation about the point indicated by
$\oplus$. In the folded sketches on the right-hand side, paths
that end on the left lead are marked with solid circles, while
those that end on the right lead are marked with open (red)
circles (as in Fig.~\ref{Fig:leftright}). When we fold all paths
into one half of the dot (see the right-hand sketches) by rotating
the other half of the dot by $180^\circ$, any path which crosses
the line between the two halves (dashed line) at $(r,p)$ reappears
at $(-r,-p)$, where $r=0$ is defined as the centre of the
inversion symmetry (marked $\oplus$). The sketch in (a) is an
enhanced forward-scattering contribution similar to that in
Fig.~\ref{Fig:leftright}(a). }
\end{figure}

The effect of inversion symmetry is similar to left-right mirror symmetry. In
particular both symmetries map the L lead onto the R lead and vice versa. As
the only major difference, inversion symmetry preserves the orientation of
closed loops, which results in a different magnetic-field dependence of the
interference contributions.

Fig.~\ref{Fig:inversion} shows the pairs of paths which give
interference corrections due to the inversion symmetry.  Here path
$\gamma'$ is the symmetric partner of path $\gamma$ if the mapping
from $\gamma'$ to $\gamma$ is inversion plus time-reversal.
Comparison with Fig.~\ref{Fig:leftright} reveals the similarity of
these pairs with the equivalent pairs in a left-right mirror
symmetric dot. The only difference is the orientation of loops.
Looking at Fig.~\ref{Fig:inversion}(a) one sees that the flux
enclosed by a path is unchanged under inversion, but then changes
sign under time-reversal.  Thus a path encloses the opposite
amount of flux from its ``symmetric partner''. This is the same as
in a weak localization loop \cite{Richter-Sieber} (where the
symmetric partner of a path is simply its time-reverse), and so we
can expect that the interference corrections induced by the
inversion symmetry follow
 the same Lorentzian
magnetic-field suppression [see Eq.~(\ref{Eq:Zwl})].

When performing the calculation of the contributions $\langle \de
g\rangle_{\rm inv}$, the only difference from that for $\langle \de
g\rangle_{\rm LR}$ in Section~\ref{sect:LR} is therefore an extra suppression
of the form $\exp[-(t/\tau_0)(B/B_0)^2]$, which is incurred during the time
when paths are approximate symmetric partners (rather than being
approximately identical). This exponential suppression can be obtained
\cite{Richter-Sieber} by noting that a path segment acquires a phase $eB{\cal
A}/h$ between two bounces from the dot's boundary, where ${\cal A}$ is the
directed area enclosed by the triangle defined by the straight path segment
and a third point (say, the point at the centre of the dot).  For a chaotic
path, ${\cal A}$ can be treated as a random variable with $\langle {\cal A}
\rangle=0$ and $\langle {\cal A}^2 \rangle\sim A^2$, where $A$ is the area of
the dot. Assuming the phase acquired between two bounces is small, we can use
the central limit theorem to obtain the average value of the phase factor
$\exp[\rmi \Phi_B(t)]$ for a path segment of duration $t$ (which therefore
bounces $t/\tau_0$ times),
\begin{eqnarray}
\Big\langle \exp[\rmi \Phi_B (t)] \Big\rangle &=&
\prod_{i=0}^{t/\tau_0} \Big\langle \exp[\rmi eB{\cal A}/h]\Big\rangle
\nonumber \\ 
&=& \exp[-(t/\tau_0)(B/B_0)^2].
\end{eqnarray}
In all contributions, this factor simply adds to the asymmetry-
and dephasing-induced suppression. The final result is
\begin{eqnarray}
\langle g\rangle_{\rm inv}
&=& {N\over 2} + {1 \over 4}  \e^{-\tEc/\tD}
\left[
Z_{\rm inv}(B,\gamma_{\rm asym},\gamma_\phi)
- Z(B,\gamma_\phi) \right]
\nonumber \\
& &\ +\ {\cal O}[N^{-1}]
\end{eqnarray}
where we define $N=N_{\rm L}=N_{\rm R}$. The suppression due to a
magnetic field, asymmetry or dephasing is given by
\begin{eqnarray}
Z_{\rm inv}(B,\gamma_{\rm asym},\gamma_\phi) = {\exp[
-\gamma_\phi\tilde{\tau} -\gamma_{\rm asym}\tilde{\tau}_{\rm asym} ]
\over 1+(B/B_{\rm c})^2 +(\gamma_{\rm
asym}+\gamma_\phi)\tD}. \label{Eq:Z_inv}
\end{eqnarray}
This is similar to Eq.~(\ref{Eq:Z_LR1}) for a left-right symmetry,
 but has an extra magnetic-field dependent
term in the denominator. Just as for left-right symmetry we have
also calculated the contributions to reflection, and verified that
particle number is conserved.

\section{Up-down mirror symmetry}
\label{sect:UD}

\begin{figure}
\centerline{\hbox{\includegraphics[width=1.0\columnwidth]{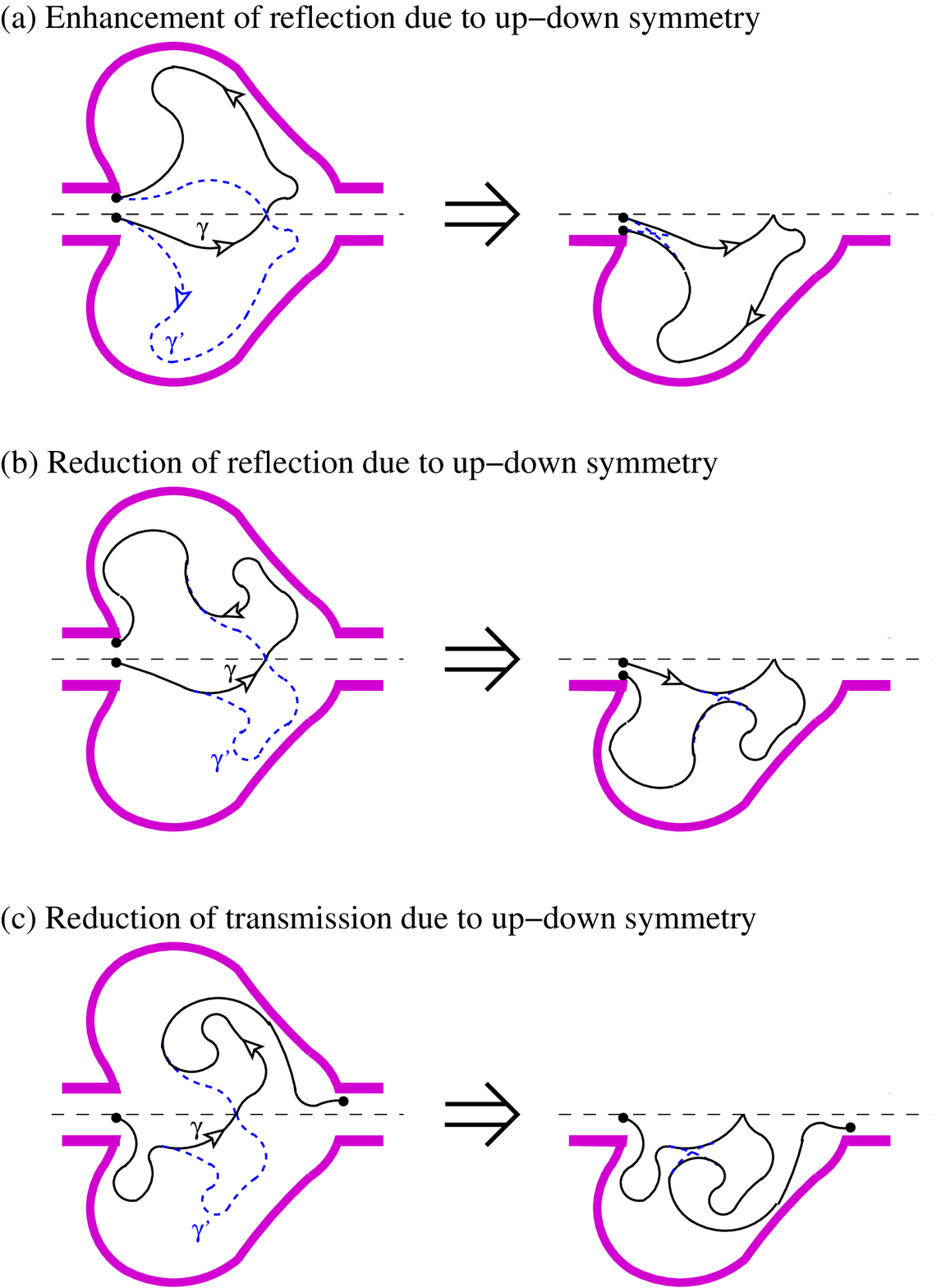}}}
\caption[]{\label{Fig:updown} (colour online). 
Contributions to weak localization
due to an up-down mirror symmetry for a dot with symmetric leads.
The sketch in (a) is an enhanced backscattering contribution;
there are no enhanced forward-scattering contributions (unlike for
left-right and inversion-symmetric dots). }
\end{figure}

The case of up-down mirror symmetry is different from the
symmetries we have already considered because this symmetry maps
each lead onto itself. Figure \ref{Fig:updown} shows the pairs of
paths which contribute to the associated interference corrections;
in contrast to the earlier cases there is now only one
contribution to transmission [that shown in
Fig.~\ref{Fig:updown}(c)].

Despite the difference in topology from the equivalent pairs in a
left-right symmetric dot (shown in Fig.~\ref{Fig:leftright}), the
calculations of the contributions are still similar in the two
cases. The integrals we have to evaluate for each contribution in
Fig.~\ref{Fig:updown} have the same form as the equivalent
contribution in Fig.~\ref{Fig:leftright}, except that some
integrals over lead R now become integrals over  lead L,  and
vice-versa. For symmetric leads (i.e., leads that may have
different widths, but must both be centred on the symmetry axis)
we then find
\begin{eqnarray}
\langle \delta g\rangle_{\rm UD} 
&=& -{N_{\rm L}N_{\rm R} \over
(N_{\rm L}+N_{\rm R})^2 }\exp[-\tEc/\tD^{\rm (UD)}] 
\nonumber \\
& & \qquad \times Z_{\rm
UD}(\gamma_{\rm asym},\gamma_\phi),  \label{Eq:gUD}
\end{eqnarray}
where $Z_{\rm UD}(\gamma_{\rm asym},\gamma_\phi)$ has the same
form as $Z_{\rm LR}(\gamma_{\rm asym},\gamma_\phi)$ given in
Eq.~(\ref{Eq:Z_LR1}). For the same reason as for left-right
symmetry (but unlike the case of inversion symmetry), there is no
magnetic-field dependence. If we sum all the contributions in
Fig.~\ref{Fig:updown}  we get zero, so the above reduction of
transmission is exactly compensated by an enhancement of
reflection, and therefore particles are conserved.

In combination with the weak localization correction, the average
conductance of an up-down mirror-symmetric dot is
\begin{eqnarray}
\langle g\rangle_{\rm UD}
&=& {N_{\rm L}N_{\rm R} \over N_{\rm L}+N_{\rm R}}
\\
& &- {N_{\rm L}N_{\rm R} \over (N_{\rm L}+N_{\rm R})^2}
\Bigg[
\exp[{-\tEc/\tD^{\rm (UD)}}] \ Z_{\rm UD} (\gamma_{\rm asym},\gamma_\phi)
\nonumber \\
& & \qquad \qquad + \exp[{-\tEc/\tD}] \ Z
(B,\gamma_\phi) \Bigg] 
\ + {\cal O}[N_{\rm L,R}^{-1}],
\nonumber 
\end{eqnarray}
where for simplicity we assume that $\tilde{\tau}$ and
$\tilde{\tau}_{\rm asym}$ are small enough to be neglected. For
perfect symmetry and vanishing magnetic field, the interference
correction to the conductance is exactly twice that of weak
localization alone; this is the case for any lead widths,
Ehrenfest time or dephasing rate.

\section{Four-fold symmetry}
\label{sect:4fold} A quantum dot with four-fold symmetry has all
three of the spatial symmetries that we discuss in this article.
The interference corrections to the conductance of such a
four-fold symmetric quantum dot are simply the sum of the
corrections due to each of these three symmetries. The presence of
the extra symmetries has no effect on the contributions which do
not respect those symmetries, so we can directly take the results
we already calculated. The only problem is that we have to avoid
double-counting any contributions where a path and its symmetric
partner respect more than one of the above symmetries. Luckily
such path-pairs are vanishingly rare. For example, to construct a
pair of paths related by both left-right and inversion symmetry,
one require that both paths go through the left-right mirror axis
exactly perpendicular to that axis; only a vanishing small
proportion of the paths in the dot do this. Thus
\begin{eqnarray}
\label{Eq:g4F}
\langle \de g\rangle_{\rm 4F}
&=& \langle \de g\rangle_{\rm LR} + \langle \de g\rangle_{\rm UD}
+\langle \de g\rangle_{\rm inv}
\\
&=& {1\over 4} \exp[-\tEc/\tD]
\, \big[Z_{\rm LR} (\gamma_{\rm asym}^{\rm LR},\gamma_\phi)
\nonumber \\
& & +Z_{\rm inv} (B,\gamma_{\rm asym}^{\rm inv},\gamma_\phi)
-Z_{\rm  UD} (\gamma_{\rm asym}^{\rm UD},\gamma_\phi) \big].\quad \quad 
\end{eqnarray}
For an arbitrary deformation of the dot, one can assume that all
three symmetries will be affected about equally,  $\gamma_{\rm
asym}^{\rm LR} \simeq \gamma_{\rm asym}^{\rm inv} \simeq
\gamma_{\rm asym}^{\rm UD}$. However, one could also consider a
deformation which respects one of the symmetries, so that the
corresponding $\gamma_{\rm asym}$ will be zero, but the other two
will be finite (and typically of similar magnitude). For perfect
symmetry, we have $\langle \de g\rangle_{\rm LR} = -\langle \de
g\rangle_{\rm UD}$. Hence in that case $\langle \de g\rangle_{\rm
4F}=\langle \de g\rangle_{\rm inv}$ for arbitrary magnetic fields,
Ehrenfest time and dephasing rate.

\begin{figure*}
\centerline{\hbox{\includegraphics[width=0.68\textwidth]{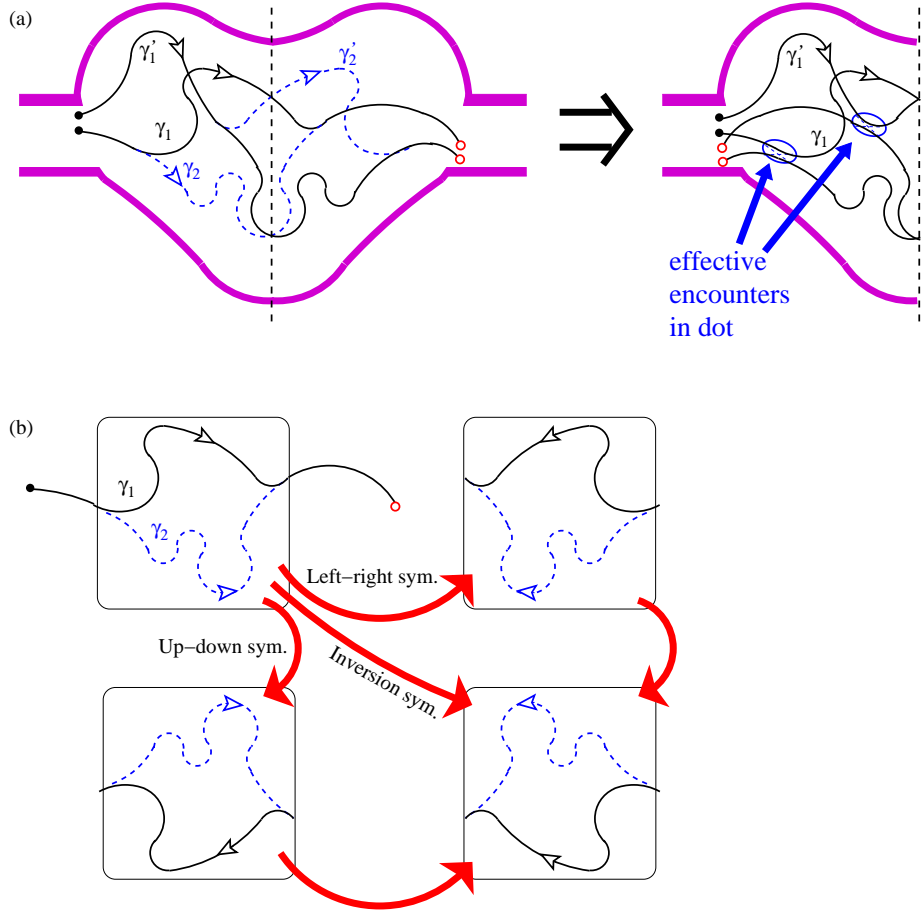}}}
\caption[]{\label{Fig:ucfs} (colour online). 
(a) A contribution to universal
conductance fluctuations in a left-right mirror-symmetric system.
On the right we show the same contribution folded into half the dot,
such that one can see the two effective encounters
(we only show the dashed paths at the encounters).
(b) The full list of pairings for path $\gamma_1$ and $\gamma_2$,
with left-right, up-down and inversion symmetry. There are also
contributions in which paths are paired with the time-reverse of their images
(arrows reversed); indeed (a) is an example of this for a left-right
symmetric system.}
\end{figure*}

\section{Universal conductance fluctuations}
\label{sect:UCFs}

In the case of perfect spatial symmetry the universal conductance
fluctuations (UCFs) are much easier to analyze than the
interference corrections to the conductance itself. Each spatial
symmetry doubles the size of every contribution, in exactly the
way that introducing time-reversal symmetry doubles the
conductance fluctuations in a dot without spatial symmetry (this
is also the case in disordered systems, for a review see
Ref.~\cite{Montambaux-book}). This leads immediately to the
results given in Table~\ref{table1}.

To see from  semiclassics why this is the case, we turn to
Fig.~\ref{Fig:ucfs}.  Contributions to conductance fluctuations
involve two paths, $\gamma_1$ and $\gamma_2$, from $y_0$ on lead L
to $y$ on lead R, which are completely uncorrelated for part (or
all) of their time in the dot. These paths are then paired with
another two paths, $\gamma'_1$ and $\gamma'_2$ (this time from
$y'_0$ on lead L to $y'$ on  lead R). If there are no spatial
symmetries and no time-reversal symmetry, then there is only one
possible pairing which survives averaging: the parts of $\gamma_1$
and $\gamma_2$ which are not paired with each other must be paired
with $\gamma'_2$ and $\gamma'_1$ respectively. Time-reversal
symmetry induces a second possibility, namely to pair $\gamma'_2$
and $\gamma'_1$ with the {\it time-reverse} of the parts of
$\gamma_1$ and $\gamma_2$ which are not paired with each other.
Therefore UCFs in time-reversal symmetric systems are double the
magnitude of UCFs in systems without time-reversal symmetry.

The same logic applies to systems with spatial symmetries. If the
system has a left-right symmetry, then one can pair $\gamma'_2$
and $\gamma'_1$ with the {\it mirror image} of the parts of
$\gamma_1$ and $\gamma_2$ which are not paired with each other
[see Fig.~\ref{Fig:ucfs}(a) for an example of this]. Thus the
presence of a left-right symmetry doubles the magnitude of the
UCFs.  The same additional pairings also occur for up-down and
inversion symmetry. For  multiple symmetries the effects are
multiplicative; assuming that there are $n$ independent symmetries
one finds $2^n$ ways to pair paths, meaning the UCFs are $2^n$
times larger than for a system with no symmetries.  A four-fold
symmetry has two independent symmetries (one possible choice is
left-right and up-down mirror symmetry, the inversion symmetry
being then just a consequence of these two).
Fig.~\ref{Fig:ucfs}(b) shows the four possible pairings in such a
case.

\subsection{The effect of magnetic fields and asymmetries on UCFs}

Asymmetries have a similar effect on paths paired with their
mirror-symmetric partner, as magnetic fields have on paths paired
with their time-reverse. Path-pairs where each path is the
time-reverse of the other (cooperons) decay with a factor of
$\exp[-t/\tau_{\rm D} - \tau_0^{-1}(B/B_0)^2t]$, rather than the
factor of $\exp[-t/\tau_{\rm D}]$ for pairs where both paths are
the same (diffusons). Upon integrating over the length $t$ of such
pairs,  one sees that for each pair in which the paths are the
time-reverse of each other, one gets an extra factor of
$[1+(B/B_{\rm c})^2]^{-1}$ where $B_{\rm c}$ is given in Eq.\
(\ref{eq:bc}). Ref.~\cite{Brouwer-Rahav-ucfs2} shows that there
are two such pairs for zero Ehrenfest time, but only one pair for
infinite Ehrenfest time; this observation leads to the result
given in Eq.~(\ref{Eq:UCF-suppression}).

This methodology can be directly carried over to asymmetries in a
spatially symmetric system. Just as for weak localization, pairs
where each path is the mirror-symmetric of the other decay with a
factor of $\exp[-t/\tau_{\rm D} - \gamma_{\rm asym} t]$ rather
than simply $\exp[-t/\tau_{\rm D}]$. The asymmetry-induced decay
rate $\gamma_{\rm asym}$ is given by
Eq.~(\ref{Eq:gamma-asym-gentle-ripple}) for asymmetries of the type
shown in Fig.~\ref{Fig:asymm}(a), and  by
Eq.~(\ref{Eq:gamma-asym-local-deform}) for asymmetries of the type
shown in Fig.~\ref{Fig:asymm}(b).

We next consider the effect of perpendicular magnetic fields. Looking at the
left-right mirror-symmetric paths in Fig.~\ref{Fig:ucfs}(b), we see that they
enclose the opposite directed area, so if we pair segments of path
$\gamma'_2$ and $\gamma'_1$ with the mirror image of segments of $\gamma_1$
and $\gamma_2$, respectively, then the Aharonov-Bohm flux will be twice that
enclosed by paths $\gamma_1$ and $\gamma_2$. In contrast, if we pair segments
of path $\gamma'_2$ and $\gamma'_1$ with the {\it time-reverse} of the mirror
image of segments of $\gamma_1$ and $\gamma_2$, respectively, then the
Aharonov-Bohm flux will cancel. Thus we have a factor of $[1+ (B/B_{\rm c})^2
+ \gamma_{\rm asym}\tau_{\rm D}]^{-1}$ for each pair containing paths that
are the mirror image of each other, and a factor of $[1+ \gamma_{\rm
asym}\tau_{\rm D}]^{-1}$ for each pair containing paths where one is the
time-reverse of the mirror image of the other.

As for the contributions in  Ref.~\cite{Brouwer-Rahav-ucfs2}, the
zero-Ehrenfest time contributions have two pairs which are not the
same, while the infinite-Ehrenfest time contributions have only
one. Thus for left-right mirror symmetry we find
\begin{eqnarray}
{\rm var}[g]_{\rm LR} = {1 \over 16} \big[Z^{\rm ucf}(B) + Z^{\rm
ucf}_{\rm sym}(B,\gamma_{\rm asym}^{\rm LR})\big],
\end{eqnarray}
where $Z^{\rm UCF}(B)$ is given in Eq.~(\ref{Eq:UCF-suppression}) and
\begin{eqnarray}
& & \hskip -10mm 
Z^{\rm ucf}_{\rm sym}(B,\gamma_{\rm asym})
\nonumber \\
&=& {\e^{-\tEo/\tau_{\rm D}}
\over [1+\gamma_{\rm asym}\tau_{\rm D}]^2}
+{\e^{-\tEo/\tau_{\rm D}}
\over [1+\gamma_{\rm asym}\tau_{\rm D} +(B/B_{\rm c})^2]^2}
\nonumber \\
& & + {1-\e^{-\tEo/\tau_{\rm D}} \over 1+\gamma_{\rm
asym}\tau_{\rm D}} + {1-\e^{-\tEo/\tau_{\rm D}} \over
1+\gamma_{\rm asym}\tau_{\rm D}+(B/B_{\rm c})^2} .
\label{Eq:Zucf-sym}
\end{eqnarray}
For perfect mirror symmetry (as in the absence of spatial
symmetries \cite{Brouwer-Rahav-ucfs2}), the magnitude of the UCFs
for $B \ll B_{\rm c}$ and $B \gg B_{\rm c}$ is independent of the
Ehrenfest time,  but the crossover from one to the other is given
by a Lorentzian in the classical limit ($\tEo \gg \tau_{\rm D}$)
and by a Lorentzian-squared in the random-matrix limit ($\tEo \ll
\tau_{\rm D}$).  Similarly for  $B \ll B_{\rm c}$ or $B \gg B_{\rm
c}$, the crossover from symmetric to asymmetric is given by
$[1+\gamma_{\rm asym}\tau_{\rm D}]^{-1}$ in the classical limit
and by $[1+\gamma_{\rm asym}\tau_{\rm D}]^{-2}$ in the
random-matrix limit.

The same logic follows for up-down symmetry. In contrast, for the
case of an inversion symmetry there is a small difference. Then,
if we pair segments of path  $\gamma'_2$ and $\gamma'_1$ with the
image (under the symmetry) of segments of $\gamma_1$ and
$\gamma_2$, respectively,  the  Aharonov-Bohm flux cancel. If on
the other hand we pair segments of path  $\gamma'_2$ and
$\gamma'_1$ with the {\it time-reverse} of the image (under the
symmetry) of segments of $\gamma_1$ and $\gamma_2$, respectively,
then the Aharonov-Bohm will be twice that enclosed by paths
$\gamma_1$ and $\gamma_2$. This means that the contributions to
UCFs which survive at finite magnetic field ($B\gg B_{\rm c}$) for
inversion symmetry are exactly those that do not survive for
left-right or up-down mirror symmetry, and vice versa. Even though
different individual contributions are suppressed for the
different symmetries, the total suppression of UCFs by magnetic
fields or asymmetries has the same form in all three cases, given
by $Z^{\rm ucf}_{\rm sym}(B,\gamma_{\rm asym})$ in
Eq.~(\ref{Eq:Zucf-sym}), where $\gamma_{\rm asym}$ measures the
deviations from the relevant symmetry.

From this one immediately gets
\begin{eqnarray}
{\rm var}[g]_{\rm inv} = {1 \over 16}
\big[Z^{\rm ucf}(B) + Z^{\rm ucf}_{\rm sym}(B,\gamma_{\rm asym}^{\rm inv})\big]
\end{eqnarray}
and
\begin{eqnarray}
{\rm var}[g]_{\rm UD} = {N_{\rm L}^2 N_{\rm R}^2 \over (N_{\rm L}
+N_{\rm R})^4 } \big[Z^{\rm ucf}(B) + Z^{\rm ucf}_{\rm
sym}(B,\gamma_{\rm asym}^{\rm UD})\big].\ \ 
\end{eqnarray}

For four-fold symmetry, the expression for the UCFs contains four
terms; the first is from Eq.~(\ref{Eq:UCF-suppression}) while each
of the others comes from one of the spatial symmetries that a
four-fold system respects. Thus
\begin{eqnarray}
{\rm var}[g]_{\rm 4F} &=& {1 \over 16}
\big[Z^{\rm ucf}(B)
+ Z^{\rm ucf}_{\rm sym}(B,\gamma_{\rm asym}^{\rm LR})
\nonumber \\
& &
+ Z^{\rm ucf}_{\rm sym}(B,\gamma_{\rm asym}^{\rm inv})
+ Z^{\rm ucf}_{\rm sym}(B,\gamma_{\rm asym}^{\rm UD})\big]. \quad
\end{eqnarray}
As we discussed for weak localization, an arbitrary deformation of
the dot is likely to affect all three symmetries about equally,
with $\gamma_{\rm asym}^{\rm LR} \simeq \gamma_{\rm asym}^{\rm
inv} \simeq \gamma_{\rm asym}^{\rm UD}$. However, one could also
have a deformation which respects one of the symmetries; then the
corresponding $\gamma_{\rm asym}$ will be zero, while the other
two are finite.

For perfect spatial symmetry with $B\ll B_{\rm c}$ we have $Z^{\rm
ucf}_{\rm sym}(B,\gamma_{\rm asym})=2$, and a four-fold system has
UCFs of magnitude $1/2$ irrespective of the Ehrenfest time. For
perfect symmetry with $B\gg B_{\rm c}$ we have $Z^{\rm ucf}_{\rm
sym}(B,\gamma_{\rm asym})=1$, and a four-fold system has UCFs of
magnitude $1/4$ irrespective of the Ehrenfest time.

\section{\label{sect:RMT}Comparison to random-matrix theory}

\begin{table}
\begin{tabular}{|l|c c|}
\hline \hline & $ B=0 $& $B\gg B_{\rm c}$\\  \hline
no spatial sym. &     COE($M$)        &               CUE($M$)\\
left-right  sym. & $A^\dagger$ COE$^2$($M/2$) $A$ & $A^\dagger$ COE($M$) $A$ \\
inversion sym.   &    $A^\dagger$ COE$^2$($M/2$) $A$ & $A^\dagger$
CUE$^2$($M/2$) $A$
\\
up-down  sym.& $C^T$ COE$^2$($M/2$) $C$ & COE($M$) \\
four-fold  sym.&
$C^T$[$A^\dagger$ COE$^2$($M/4$) $A$]$^2$$C$ &$A^\dagger$
COE$^2$($M/2$) $A$
\\
\hline \hline
\multicolumn{3}{|r|}{$\phantom{\begin{array}{c} 1 \\ 1 \\ 1 \end{array}}$
with 
$A=2^{-1/2} \left(\begin{array}{cc} 1 & 1 \\  i & -i \\ \end{array} \right)$
and $C_{i,j}=\delta_{2i-1,j}+\delta_{2i-M,j}$
}
\\
\hline\hline
\end{tabular}
\caption{\label{table2}
Pure random-matrix ensembles for each spatial symmetry, in absence
or presence of a magnetic field. The block composition
of two identical matrix ensembles of dimension $M$ is abbreviated
as $X^2(M)=X(M)\otimes X(M)$. We only consider the case  
$M \mbox{mod 4}=0$; in the general case one encounters  block composition
of ensembles with dimensions that differ at most by 1.}
\end{table}

\begin{figure*}
\centerline{\hbox{\includegraphics[width=0.8\textwidth]{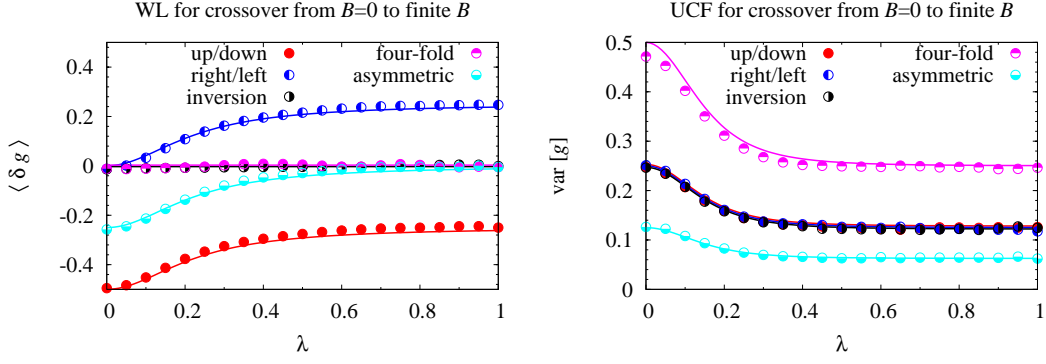}}}
\caption[]{\label{Fig:numerics1} (colour online). 
Effect of a magnetic field
$B=(\lambda/\lambda_{\rm c}) B_{\rm c}\simeq \lambda B_0$ on the
weak-leak localization correction (left panels) and the
conductance fluctuations (right panels) for quantum dots of
different spatial symmetry. The solid lines show the universal
crossover functions (\ref{dglambda}) (\ref{varglambda}),  derived
in this paper based on semiclassical arguments. The data points
show the result of numerical computations based on the
interpolation formulae Eqs.\ (\ref{eq:interf}) and (\ref{eq:dyns})
between the random-matrix ensembles tabulated in Tab.
\ref{table2}. In these computations, $M=1000$ and $N=50$,
resulting in $\lambda_{\rm c}=1/\sqrt{20}$. Each data point is
obtained by averaging over 5000 realizations.
 }
\end{figure*}

\begin{figure*}
\centerline{\hbox{\includegraphics[width=0.8\textwidth]{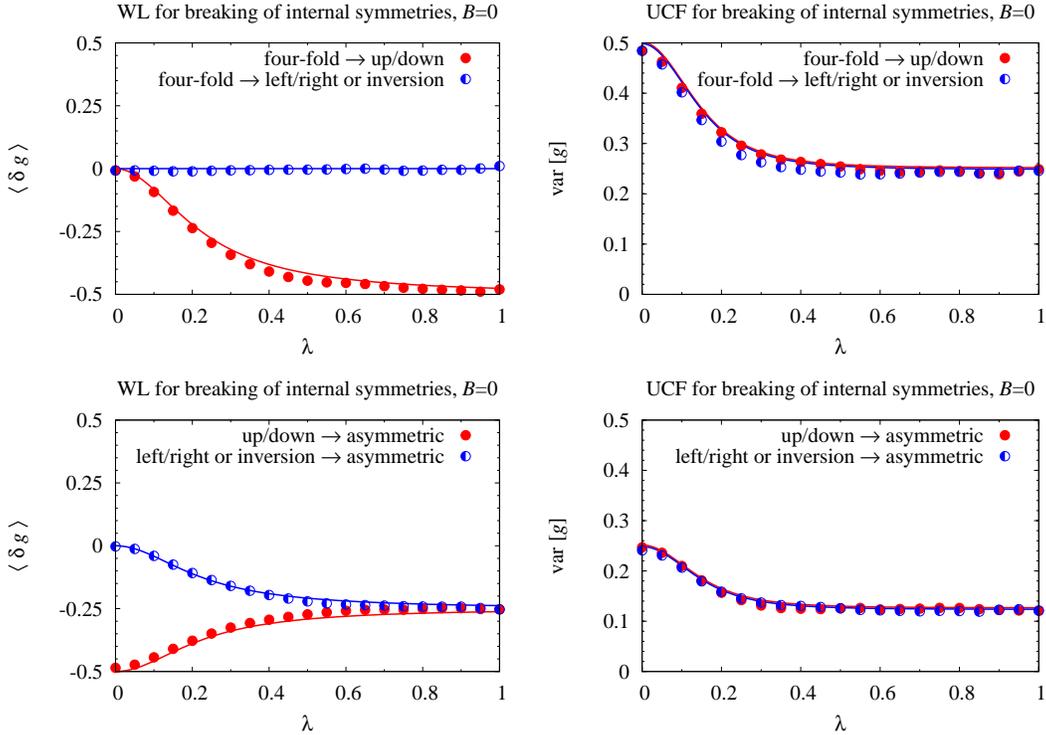}}}
\caption[]{\label{Fig:numerics2} (colour online). 
Same as Fig.\
\ref{Fig:numerics1}, but for transitions induced by the breaking
of internal spatial symmetries ($B=0$).
 }
\end{figure*}

\begin{figure*}
\centerline{\hbox{\includegraphics[width=0.8\textwidth]{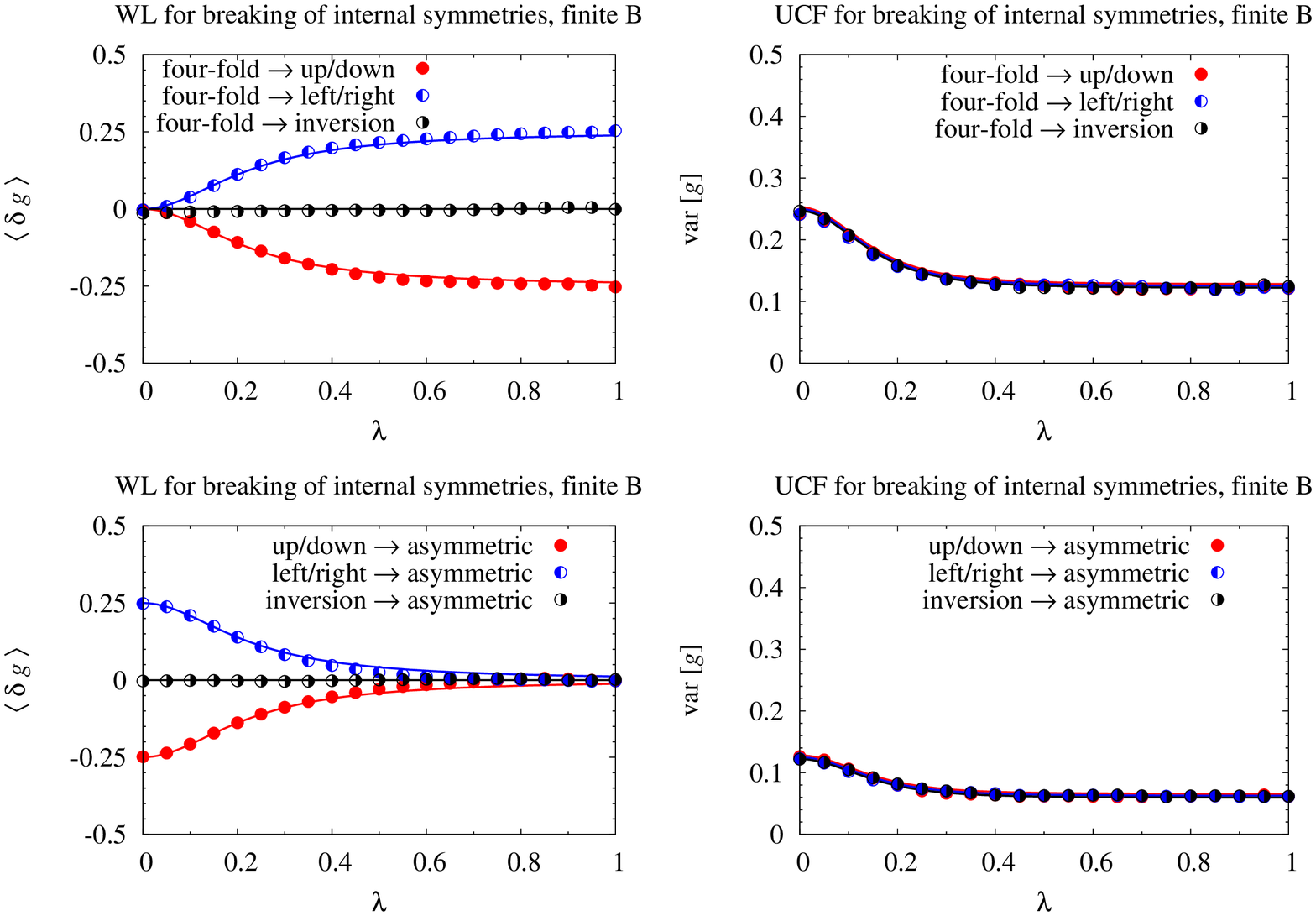}}}
\caption[]{\label{Fig:numerics3} (colour online). 
Same as Fig.\
\ref{Fig:numerics2}, but for a finite magnetic field.
 }
\end{figure*}

In this section we compare the semiclassical predictions derived
in the previous sections to numerical results obtained from a
phenomenological random-matrix model. For simplicity, we assume
$\tEo\ll\tD$ and no dephasing and consider the crossover of
quantum-interference effects induced by a magnetic field in
cavities of fixed spatial symmetry, as well as the crossover when
the magnetic field is fixed but the shape is changed to break one
or more spatial symmetries. We concentrate on the case $N_{\rm
L}=N_{\rm R}= N$.

Our results are based on an efficient interpolation scheme between `pure'
random-matrix ensembles which represent the limiting scenarios of vanishing
or strong magnetic field at fully preserved or broken spatial symmetry. We
interpolate between these ensembles by means of a procedure which combines
the ideas of two constructions used in earlier related works: the stub model
of Refs.\ \cite{Brouwer-Frahm-Beenakker,Brouwer-Cremers-Halperin}, and the
dynamical scattering model of Refs.\
\cite{Jacquod-Schomerus-Beenakker,Tworzydlo-Tajic-Schomerus-Beenakker}.

The main common feature of our interpolation procedure and the
stub model is the idea to introduce a controllable coupling to a
symmetry-breaking or magnetic auxiliary system. Ordinarily, this
coupling is introduced on the basis of an internal scattering
matrix which partially couples to the leads and partially couples
to the auxiliary system.

It is beneficial to adapt this approach in a manner which more
clearly separates the modifications in the internal dynamics from
the coupling to the leads. This can be achieved by first focussing
on an $M\times M$-dimensional unitary evolution operator $F$ which
represents the internal dynamics of the system. The operator $F$
is taken from a crossover ensemble which we construct by coupling
the symmetric system to a symmetry-breaking system, where the flux
between the systems is controlled by a parameter (denoted by
$\lambda$). The result of this construction is the following
simple interpolation formula of the internal operator (given in
two equivalent forms):
\begin{eqnarray}
F(\lambda)&=&(F_0 + \lambda F_1)  (1+\lambda F_1^{-1} F_0)^{-1}\\
&=& (F_0 + \lambda F_1) (F_1+\lambda F_0)^{-1} F_1.
\label{eq:interf}
\end{eqnarray}
This is a unitary matrix which smoothly interpolates between
$F(\lambda=0)=F_0$ and $F(\lambda=1)=F_1$.

For reference, Table \ref{table2} lists the pure random-matrix
ensembles for the internal operator $F$ in a form in which the
leads specified below respect the spatial symmetries (see the
second part, Ref.\ \cite{whitney-schomerus-kopp-2nd}, for a
generalized model which allows to displace the leads from their
symmetric positions). As explained in
Appendix \ref{app:A}, the form of
these matrix ensembles is closely related to the presence or
absence of generalized time-reversal symmetry. Drawing pairs
$F_0$, $F_1$ of matrices  at random from two different ensembles
listed in Table \ref{table2}
 delivers a smooth interpolation of the internal dynamics
between specified symmetry scenarios.

The phase-coherent transport through the open system can be
obtained from the $2N\times 2N$-dimensional scattering matrix $S=\left(%
\begin{array}{cc}
  r & t' \\
  t & r' \\
\end{array}%
\right)$, with blocks containing reflection amplitudes $r$, $r'$,
and transmission amplitudes $t$, $t'$. The dimensionless
conductance follows from the Landauer formula $g=\tr t^\dagger t$
\cite{Beenakker-1997}. Building up on the internal dynamics, we
specify the scattering matrix as
\cite{Jacquod-Schomerus-Beenakker,Tworzydlo-Tajic-Schomerus-Beenakker}
\begin{equation}
S= P^T (1-FQ)^{-1} F P, \label{eq:dyns}
\end{equation}
where \begin{equation}
 P=\left(%
\begin{array}{cc}
  1_{N\times N} & 0_{N\times N} \\
  0_{M/2-N\times N} & 0_{M/2-N\times N} \\
  0_{N\times N} & 1_{N\times N} \\
  0_{M/2-N\times N} & 0_{M/2-N\times N} \\
\end{array}%
\right) \label{eq:leadprojector}
\end{equation}
is a matrix of dimensions $M\times 2N$ which projects onto the
leads, while $Q=1-PP^T$ projects onto the hard walls where no
leakage occurs. Equation (\ref{eq:dyns}) lifts the interpolation
(\ref{eq:interf}) of the internal dynamics to an interpolation
between different scattering matrix ensembles.

While the resulting  expressions for the interpolated scattering
matrix are quite similar to the stub model, our construction
possesses some practical advantages. In particular, the stub model
has the deficit that symmetry is not broken for trajectories which
avoid the stub (corresponding to a direct transfer from entrance
to exit). Total crossover in the stub model therefore can only be
achieved in the limit $M\gg N$. That this constraint does not
present itself in our procedure is convenient not only because $M$
and $N$ carry precise microscopic meaning, but also because the
ratio $M/2N=\tau_D/\tau_0$ represents the dimensionless dwell time
featuring in the semiclassical crossover formulae. At the same
time, our interpolation procedure avoids the computational
overhead which one would encounter by basing the crossover on
other microscopic schemes, such as the Pandey-Mehta class of
Hamiltonians \cite{Beenakker-1997}.

In order to determine the physical meaning of the phenomenological
parameter $\lambda$ we consider the universal limit $M\gg N \gg
1$, in which a particle typically undergoes many internal
reflections before leaving the system. Each internal reflection
probes the broken symmetry, so that we expect that the crossover
is essentially completed for small values of $\lambda$, on a
crossover scale
\begin{equation}
\sqrt{\tau_0/2\tau_D}\simeq \sqrt{N/M} \equiv \lambda_{\rm c}
\label{eq:lc}
\end{equation}
which scales with the inverse square-root of the dimensionless
dwell time.
 Comparing
this crossover acceleration with the effect of a magnetic field in
microscopic models (Ref.\ \cite{Beenakker-1997}, as well as the
semiclassical theory employed in the present paper) allows us to
relate the interpolation parameter $\lambda$ to the magnetic field
through the quantum dot,
\begin{equation}
\lambda = (B/B_{\rm c})\lambda_{\rm c}\simeq B/B_0,
\end{equation}
where we used Eq.\ (\ref{eq:bc}).
 By construction, the  crossover scale (\ref{eq:lc}) also
applies to interpolation between ensembles of different spatial symmetry.

The semiclassical considerations in this paper predict that the
crossover of the weak localization correction and the conductance
fluctuations obey a universal functional dependence on
$\lambda/\lambda_{\rm c}$ for all types of symmetry breaking,
which for $M\gg N$ ($\lambda_{\rm c}\ll 1$) can be written as
\begin{equation}
\delta g(\lambda)= \delta g(1)+ \frac{\delta g(0)-\delta
g(1)}{1+(\lambda/\lambda_{\rm c})^2}, \label{dglambda}
\end{equation}
\begin{equation}
{\rm var} g(\lambda)= {\rm var}  g(1)+ \frac{{\rm var}  g(0)-{\rm
var}
 g(1)}{[1+(\lambda/\lambda_{\rm c})^2]^2}.  \label{varglambda}
\end{equation}
Here $g(0)$ and $g(1)$ refer to the random-matrix values given in
Tab.\ \ref{table1}.

Figures \ref{Fig:numerics1}, \ref{Fig:numerics2}, and
\ref{Fig:numerics3} compare these analytical predictions to the
results of  numerical computations for the most important symmetry
breaking scenarios. In each figure, the data points are obtained
from averaging over 5000 realizations of the corresponding
interpolation ensemble. The internal matrix dimension $M=1000$ and
number of channels $N=50$ translates into a critical value
$\lambda_{\rm c}=1/\sqrt{20}$ of the interpolation parameter. The
solid curves show our analytical predictions, Eqs.\
(\ref{dglambda}) and (\ref{varglambda}). Figure
\ref{Fig:numerics1} benchmarks the RMT model for the well-studied
effects of a magnetic field, demonstrating excellent agreement
with a Lorentzian crossover  for the weak localization correction
and a squared Lorentzian crossover for the universal conductance
fluctuations. Figure \ref{Fig:numerics2} presents results for
breaking of internal symmetries at vanishing magnetic field, while
Fig.\ \ref{Fig:numerics3} shows the same transitions for a finite
magnetic field. For all symmetry-breaking scenarios the numerical
results of the phenomenological random-matrix model are in
excellent agreement with the semiclassical predictions.

\section{Towards experimental observation}\label{sect:expt}

We have investigated the effect of spatial symmetries on transport
through chaotic quantum dots, putting significant effort into
studying how asymmetries cause a crossover in the weak
localization and universal conductance fluctuations (from perfect
spatial symmetry to completely broken symmetry). Understanding
this crossover is crucial for observing  symmetry-induced effects
in experiments.  
Experimenters who believe that their systems
are spatially symmetric can demonstrate it by confirming that the
interference effects have the deformation dependence that we
predict.
We are unaware of any experiments in which such symmetry effects have thus far been observed,
so here we estimates the conditions under which such 
observations would be feasible. 

The theory presented in this article is for
electron transport through quantum dots. 
Ultra-clean quantum dots are made by placing top gates on 
two-dimensional electron gases (2DEGs), where
typically the electron's wavelength $\lambda_{\rm F}\sim 50$nm \cite{best-ultraclean-samples}.  
To observe effects induced by a spatial symmetry,
 the dot's symmetry must be respected on a scale much less than 50nm.
For example, if a typical electron bounces fifty times before escaping, 
$\tD/\tau_0 = 50$, then Eqs.~(\ref{Eq:gamma-asym-gentle-ripple},\ref{Eq:Z_LR1}) indicate that symmetry induced effects will only be significant if imperfections in the dot's boundary are on a scale $\lambda_{\rm F}/\sqrt{50} \sim 7$nm.
This is a substantial experimental challenge, but it is not beyond the realms of possibility.

Note that we have neglected disorder throughout (although it can be included without difficulty as another asymmetry).  
For quantum dots this is justifiable, due to the remarkable
progress in making clean 2DEGs which now have mean-free paths between subsequent impurity scatterings of order of half a 
millimetre~\cite{best-ultraclean-samples}.  
Consider a dot with 10 modes on each lead,
meaning a lead width of $5\lambda_{\rm F} \simeq 250$nm, with 
$\tD/\tau_0 = 50$. Such a dot has $L\sim 2\mu$m, so the typical distance travelled by an electron
between entering and leaving the dot is $v_{\rm F}\tau_{\rm D} \sim 100\mu$m.
The probability to escape without hitting an impurity is thus about $80\%$,
so we can expect that disorder only reduces the effects discussed here by about $20\%$.
If, in contrast, one takes a dot with 5 modes on each lead and $\tD/\tau_0 = 30$
(then $L \sim 0.6\mu$m),
one finds a probability to escape without hitting an impurity  as high as $96\%$.

While our theory is for electrons in quantum dots, the results 
are applicable to any waves in chaotic shaped systems; 
microwaves in a cavity, ripples on water in a container, vibrations of metal plates, etc.
Turning to microwaves in chaotic-shaped cavities \cite{microwave-expt}, 
we note that the dimensionless conductances that we calculate
in this work can be interpreted as a measure of  the microwave power transmitted 
from one lead to the other. 
Experimentally, the microwave wavelengths are typically a few centimetres,
so one only needs to make a cavity within a few millimetres of perfectly symmetric to
see a significant effect.  This is easily attainable with conventional precision engineering.
However we note that our results are for leads carrying many modes, 
when most experiments to-date on microwave cavities use single-mode leads.  
In the single-mode case, the interference effects
are significantly more complicated than those described here.

\section{Conclusions}

By working with a microscopic theory (rather than the essentially
phenomenological RMT) we are not only able to extract the form of
the symmetric/asymmetric crossover, we are also able to say how
the crossover parameter $\gamma_{\rm asym}$ depends on the nature
of the deformation of the dot.  
Such knowledge is crucial in determining if an experimental system
is close to being spatially symmetric.
We find a remarkable wide
variation in this dependence, which for global deformations takes
the form of Eq.~(\ref{Eq:gamma-asym-gentle-ripple}) while for
local deformations it is given by
Eq.~(\ref{Eq:gamma-asym-local-deform}). We hope that the
semiclassical method developed here will lead to an improved
understanding of situations where RMT is hard to apply, such as
systems with a mixed phase space for which  at present only
phenomenological descriptions are available.

In the second of this pair of articles
\cite{whitney-schomerus-kopp-2nd}, we turn to the problem of a
spatially symmetric dot coupled to leads which do not respect the
symmetry.

\section{Acknowledgements}
RW thanks P.~Brouwer, P.~Marconcini and M.~Macucci for interesting 
and useful discussions.  RW and HS are grateful for the hospitality of the
Banff International Research Station, where this work was initiated.

\appendix
\section{\label{app:A}Generalized time-reversal symmetry}

The concept of a generalized time-reversal symmetry is frequently
used  in the RMT analysis of spatially symmetric systems
\cite{haake-book}. We do not explicitly use this concept in our
semiclassical calculations but here discuss it briefly because it
informs the construction of the RMT ensembles listed in
Table~\ref{table2}.

A spatially symmetric system which is not symmetric under the
standard (momentum-reversing) time-reversal symmetry (such as one
with a magnetic field $B \gg B_{\rm c}$) may still be invariant
under the combination of time-reversal and a spatial symmetry. To
see this one should note two things; firstly, {\it all} quantum
systems are symmetric under time-reversal if one also changes the
sign of the external magnetic field; secondly, mirror symmetries
(up-down or left-right  symmetry) map a system onto itself with
the opposite sign for the external (perpendicular) magnetic field.
Thus one immediately sees that an up-down symmetric system will be
invariant under the combination of time-reversal and mirror
symmetry; this is an example of a generalized time-reversal
symmetry. Combining this with the fact that odd (even) transverse
lead modes couple only to odd (even) internal states in the dot
\cite{Baranger-Mello}, we obtain the RMT ensembles for up-down
symmetry in Table \ref{table2}
--- odd (even) dot states being those that do (do not) change sign under
the up-down mapping. In particular, the generalized time-reversal
symmetry manifests itself in the fact that the ensemble is given
by combinations of COE matrices for all  magnetic fields, even for
$B\gg B_{\rm c}$ where one might naively expect CUE matrices.

The case of a left-right symmetry is a little more involved
because acting with the left-right mirror symmetry not only
changes the sign of the (perpendicular) magnetic field  but also
interchanges the leads. Thus the generalized time-reversal
symmetry here is a combination of time-reversal symmetry, the
left-right symmetry, and the re-labelling of the leads
(L$\leftrightarrow$ R) To obtain the ensembles in
Table~\ref{table2}, one combines this with the fact that all L
lead modes couple equally to even and odd dot states, while all R
modes have positive coupling to even dot states and negative
coupling to odd states. However, once again the presence of the
generalized time-reversal symmetry manifests itself in the fact
that the ensemble is given by COE matrices even when $B \gg B_{\rm
c}$ (see Table~\ref{table2}).

In the case of an inversion symmetry, there is no generalized time-reversal
symmetry, and thus the ensemble switches from COE matrices for $B \ll B_{\rm
c}$ to CUE matrices for $B \gg B_{\rm c}$ (see Table~\ref{table2}). The
reason for the absence of a generalized time-reversal symmetry is that the
inversion symmetry mapping does {\it not} change the sign of the magnetic
field.  Thus for $B \gg B_{\rm c}$, one cannot construct an invariance out of
the time-reversal and inversion symmetries.

Having outlined the concept of generalized time-reversal
symmetries, one sees that they are most natural in the abstract
space used for RMT. However, once we turn to the semiclassical
description of concrete systems in which we write contributions in
terms of time evolution in phase space, time-reversal symmetry and
spatial symmetries are most naturally treated independently from
each other.




\vskip 5mm
\begin{thebibliography}{10}

\bibitem{Bar91}
H. U. Baranger, D. P. DiVincenzo, R. A. Jalabert, and A. D. Stone,
Phys. Rev. B {\bf 44}, 10637 (1991).
\bibitem{Bar93}
H. U. Baranger, R. A. Jalabert, and A. D. Stone, Chaos {\bf 3},
665 (1993).

\bibitem{Larkin-Khem}
A. I. Larkin  and D. E. Khmelnitskii, Sov. Phys. Usp. {\bf 25},
185 (1982); D. E. Khmelnitskii, Physica B {\bf 126}, 235 (1984).
\bibitem{Cha86}
S. Chakravarty and A. Schmid, Phys. Rep. {\bf 140}, 193 (1986).
\bibitem{Aleiner-Larkin}
I. L. Aleiner and A. I. Larkin, Phys. Rev. B {\bf 54}, 14423
(1996).
\bibitem{Agam}
O.~Agam, I.~Aleiner, and A.~Larkin,
Phys.~Rev.~Lett.~{\bf 85}, 3153 (2000).


\bibitem{Berry}
M.V.~Berry, Proc. R. Soc. London A {\bf 400}, 229 (1985).

\bibitem{Sie01} M. Sieber and K. Richter, Phys. Scr. T {\bf 90}, 128 (2001);
 M. Sieber, J. Phys. A: Math. Gen. {\bf 35}, L613 (2002).

\bibitem{Haake-rmt}
S.~M\"uller, S.~Heusler, P.~Braun, F.~Haake, and A.~Altland,
Phys.~Rev.~Lett.~{\bf 93} 014103 (2004). S.~M\"uller, S.~Heusler,
P.~Braun, F.~Haake, and A.~Altland, Phys.~Rev.~E {\bf 72}, 046207
(2005). S.~Heusler, S.~M\"uller, A.~Altland, P.~Braun, and
F.~Haake Phys.~Rev.~Lett.~{\bf 98}, 044103 (2007).

\bibitem{Richter-Sieber}
 K. Richter and M. Sieber, Phys. Rev. Lett. {\bf 89}, 206801 (2002).
\bibitem{haake-conductance}
S. Heusler, S. M\"uller, P. Braun, and F. Haake, Phys.~Rev.~Lett.
{\bf 96}, 066804 (2006)
\bibitem{Jac06}
Ph. Jacquod and R. S. Whitney, Phys. Rev. B {\bf 73}, 195115 (2006).
\bibitem{Rahav-Brouwer-backscatter}
S. Rahav and P. W. Brouwer, Phys. Rev. Lett. {\bf 96}, 196804
(2006).

\bibitem{Brouwer-Rahav-ucfs}
P. W. Brouwer and S. Rahav, Phys.~Rev.~B {\bf 74}, 075322 (2006).
\bibitem{Brouwer-Rahav-ucfs2}
P. W. Brouwer and S. Rahav, Phys.~Rev.~B {\bf 75}, 201303(R) (2007).

\bibitem{Haake-fano}
P.~Braun, S.~Heusler, S.~M\"uller, and F.~Haake, J. Phys. A: Math.
Gen. {\bf 39}, 159 (2006)
\bibitem{wj-fano}
R. S. Whitney and Ph. Jacquod, Phys. Rev. Lett. {\bf 96}, 206804
(2006).
\bibitem{Haake-longpaper}
S. M\"uller, S. Heusler, P. Braun, and F. Haake, New J. Phys. {\bf
9}, 12 (2007).



\bibitem{Bohigas-Giannoni-Schmit}
O. Bohigas, M. J. Giannoni, and C. Schmit, Phys. Rev. Lett. \textbf{52}, 1
(1984).
\bibitem{Guhr}
T. Guhr, A. M{\"u}ller-Groeling, and H. A. Weidenm{\"u}ller, Phys.
Rep. \textbf{299}, 189 (1998).
\bibitem{Beenakker-1997}
C. W. J. Beenakker, Rev. Mod. Phys. \textbf{69}, 731 (1997).

\bibitem{Baranger-Mello}
H. U. Baranger and P. A. Mello, Phys. Rev. B {\bf 54}, R14297
(1996)
\bibitem{Gopar96}
V. A. Gopar, M. Mart\'inez, P. A. Mello, and H. U. Baranger, J.
Phys. A: Math. Gen. {\bf 29}, 881 (1996).

\bibitem{Martinez-Mello}
M. Mart\'inez and P. A. Mello, Phys. Rev. E \textbf{63}, 016205
(2000).

\bibitem{Kopp-Schomerus-Rotter}
M. Kopp, H. Schomerus, and S. Rotter, Phys. Rev. B {\bf 78},
075312 (2008).

\bibitem{Gopar-Rotter-Schomerus}
V. A. Gopar, S. Rotter, and H. Schomerus, Phys. Rev. B {\bf 73},
165308 (2006).

\bibitem{WMM}
R. S. Whitney, P. Marconcini, and M. Macucci,
Phys.~Rev.~Lett. {\bf 102}, 186802 (2009).

\bibitem{Schomerus-Jacquod}
H. Schomerus and Ph. Jacquod, J. Phys. A {\bf 38}, 10663 (2005).


\bibitem{Bro06-quasi-ucf}
A.~Altland, P. W.~Brouwer, and C.~Tian Phys. Rev. Lett. {\bf 99},
036804 (2007).
\bibitem{pjw-decoh}
C. Petitjean, Ph. Jacquod, and R. S. Whitney, JETP Lett. {\bf 86},
647 (2007).
\bibitem{wjp-decoh}
R. S. Whitney, Ph. Jacquod, and C. Petitjean, Phys. Rev. B {\bf
77}, 045315 (2008).



\bibitem{whitney-schomerus-kopp-2nd}
R. S. Whitney, H. Schomerus, and M. Kopp (part II of this series) 
preprint arXiv:0906.0892.




\bibitem{Vavilov-Larkin}
M. G. Vavilov and A. I. Larkin, Phys. Rev. B {\bf 67}, 115335
(2003).

\bibitem{Schomerus-Tworzydlo}
H. Schomerus and J. Tworzyd{\l}o, Phys. Rev. Lett. \textbf{93},
154102 (2004).

\bibitem{Whi07}
R. S. Whitney, Phys. Rev. B {\bf 75}, 235404 (2007).

\bibitem{Montambaux-book}
E. Akkermans and G. Montambaux,  {\it Mesoscopic physics of
electrons and photons} (Cambridge University Press, Cambridge,
2007).


\bibitem{Brouwer-Frahm-Beenakker}
P. W. Brouwer, K. M. Frahm, and C. W. J. Beenakker, Waves Random
Media \textbf{9}, 91 (1999).

\bibitem{Brouwer-Cremers-Halperin}
P. W. Brouwer, J. N. H. J. Cremers, and B. I. Halperin, Phys. Rev.
B {\bf 65}, 081302(R) (2002).

\bibitem{Jacquod-Schomerus-Beenakker}
Ph. Jacquod, H. Schomerus, and C. W. J. Beenakker, Phys. Rev. Lett.
\textbf{90}, 207004 (2003).

\bibitem{Tworzydlo-Tajic-Schomerus-Beenakker}
J. Tworzyd{\l}o, A. Tajic, H. Schomerus, and C. W. J. Beenakker, Phys. Rev. B
\textbf{68}, 115313 (2003).



\bibitem{best-ultraclean-samples}
I.~P.~Radu, J.~B.~Miller, C.~M.~Marcus, M.~A.~Kastner, L.~N.~Pfeiffer, 
and K.~W.~West, Science {\bf 320}, 899 (2008)

\bibitem{microwave-expt}
 H.-D.~Gr\"af, H.L.~Harney, H.~Lengeler, C.H.~Lewenkopf, C.~Rangacharyulu, A.~Richter, P.~Schardt, and H.A.~Weidenm\"uller,
Phys.~Rev.~Lett.~{\bf 69}, 1296 (1992).
B.~Dietz, T.~Friedrich, H.L.~Harney, M.~Miski-Oglu, A.~Richter, F.~Sch\"afer, J.~Verbaarschot, 
and H.A.~Weidenm\"uller,
Phys.~Rev.~Lett.~{\bf 103}, 064101  (2009).

\bibitem{haake-book} F. Haake, {\em Quantum Signatures of Chaos}, 2nd edition
    (Springer,
    Berlin, 2001).



\end{thebibliography}
\end{document}